\documentstyle[12pt]{article}

\def\beq{\begin{equation}}
\def\eeq{\end{equation}}

\def\bea{\begin{eqnarray}}
\def\eea{\end{eqnarray}}
\def\rar{\rightarrow}
\def\lrr{\leftrightarrow}
\def\ovl{\tilde}
\def\GeV{{\;\rm GeV}}
\def\MeV{{\;\rm MeV}}
\def\TeV{{\;\rm TeV}}
\def\blim{b_{\;\rm lim}}

\def\pl#1#2#3{
        {\it Phys.\ Lett.\ }{\bf #1}, #2 (19#3)}

\def\prl#1#2#3{
        {\it Phys.\ Rev.\ Lett.\ }{\bf #1}, #2 (19#3)}

\def\prep#1#2#3{
        {\it Phys.\ Rep.\ }{\bf #1}, #2 (19#3)}
\def\pr#1#2#3{
        {\it Phys.\ Rev.\ }{\bf #1}, #2 (19#3)}
\def\np#1#2#3{
        {\it Nucl.\ Phys.\ }{\bf #1}, #2 (19#3)}

\def\ib#1#2#3{
        {\it ibid.\ }{\bf #1}, #2 (19#3)}

\def\GB{\Gamma_V}
\def\MB{M_V}
\def\RP{{\cal R}_+}
\def\RM{{\cal R}_-}
\def\lz{{\cal L}_0}
\def\az{{\cal A}_0}
\def\au{{\cal A}_1}
\def\ad{{\cal A}_2}
\def\at{{\cal A}_3}
\def\aq{{\cal A}_4}

\def\cO{{\cal O}}
\def\cR{{\cal R}}

\def\cS{{\cal S}}
\def\cV{{\cal V}}
\def\pu{p_1}
\def\pd{p_2}
\def\ku{k_1}
\def\kd{k_2}
\def\kt{k_3}

\def\bq{\bar{q}}
\def\bl{\bar{l}}
\def\za{z_{\mbox{\tiny A}}}
\def\zb{z_{\mbox{\tiny B}}}
\def\xa{x_{\mbox{\tiny A}}}
\def\xb{x_{\mbox{\tiny B}}}

\def\xas{x^*_{\mbox{\tiny A}}}
\def\xbs{x^*_{\mbox{\tiny B}}}
\def\zas{z^*_{\mbox{\tiny A}}}

\def\xia{\xi_{\mbox{\tiny A}}}
\def\xib{\xi_{\mbox{\tiny B}}}
\def\gl{g_{\mbox{\tiny L}}}
\def\gr{g_{\mbox{\tiny R}}}
\def\fl{f_{\mbox{\tiny L}}}
\def\fr{f_{\mbox{\tiny R}}}
\def\gz{g_{\mbox{\tiny Z}}}
\def\xw{x_{\mbox{\tiny W}}}
\def\gw{g_{\mbox{\tiny W}}}
\def\qt{q_{\mbox{\tiny T}}}
\def\bqt{{\bf q}_{\mbox{\tiny T}}}

\begin{document}

\begin{flushright}
\mbox{
\begin{tabular}{l}
    FERMILAB-PUB-97/082-T\\
    SHEP-96/37
\end{tabular}}
\end{flushright}
\vskip 1.5cm
\begin{center}
\Large
{\bf Vector boson production in hadronic collisions} \\
\vskip 0.7cm
\large
R.K. Ellis \\
\vskip 0.1cm
{\small Theory Group, Fermi National Accelerator Laboratory,
P.O. Box 500, Batavia, IL 60510} \\
\vskip 0.3cm
D.A. Ross \\
\vskip 0.1cm
{\small Physics Department, University of Southampton, UK} \\
\vskip 0.3cm
Sini\v{s}a Veseli \\
\vskip 0.1cm
{\small Theory Group, Fermi National Accelerator Laboratory,
P.O. Box 500, Batavia, IL 60510} \\
\today
\end{center}
\thispagestyle{empty}
\vskip 0.7cm

\begin{abstract}
We consider the production of $\gamma^*, W$ and $Z$ vector bosons in
hadron-hadron collisions in perturbative QCD. We present results from
a new numerical program which gives a full description of the production
of the vector bosons and of their decay products. At small $\qt$ the
calculation includes resummation of large logarithms and non-perturbative
effects. The resummation is matched with the full $O(\alpha_S)$ calculation.
In addition, the program correctly reproduces the known $O(\alpha_S)$ cross
section when integrated over $\qt$. Besides presenting results for $W$ and
$Z$ production at the Tevatron, we also review constraints on the
non-perturbative functions using fixed target data on lepton pair
production, and make several observations on this topic.
\end{abstract}
\newpage

\section{Introduction}

The aim of this paper is to give a complete description of the leptons
coming from the decay of a vector boson produced in hadron-hadron
collisions. In addition to the intrinsic interest of obtaining a
complete description of vector boson production, a
precise measurement of the $W$ mass at a hadronic collider will depend
on accurate theoretical information about its production properties.
Since it is the leptonic decay products of the vector bosons which are
actually observed, it is important to include the decay so that
experimental cuts can be implemented.  Leptons produced at large
transverse momenta can come from vector bosons having either large or
small $\qt$, where $\qt$ is the transverse momentum of the vector
boson.  It is therefore essential to use a formalism which is valid for both
large and small transverse momentum of the vector bosons.

The production of vector bosons at large and small $\qt $ has been
extensively considered before.  Vector bosons at large $\qt$ attracted
early interest because they provided discrimination between the naive
parton model (which predicts a limitation on $\qt$) and an underlying field
theory (which predicts events at large $\qt$).
The transverse momentum of a vector boson recoiling against one parton was
calculated in Ref.~\cite{FM}.  The $O(\alpha_S^2)$ calculation of vector
boson production at large $\qt $ was initiated in Ref.~\cite{EMP} and
completed in Refs.~\cite{AR} and \cite{GPW}.  These papers on the
$\qt$ distribution give no information on the distribution of the leptons
into which the vector bosons decayed. This deficiency was remedied in
Ref.~\cite{AL} for the $O(\alpha_S)$ case and in Refs.~\cite{M,GGK} for the
$O(\alpha_S^2)$ case.

However the bulk of the data is not at large $\qt $. At small $\qt $
order by order in perturbation theory we encounter large logarithms of
$Q^2/\qt^2$, where $Q$ is the vector boson mass.
For example, at small $\qt$ the leading term in the cross
section is of the form
\begin{equation}
\frac{d \sigma}{d \qt ^2} \sim  \frac{\alpha_S}{\qt ^2}  \ln \frac{Q^2}{\qt
^2}\ .
\end{equation}
The logarithms can be resummed to give a Sudakov form
factor \cite{DDT,PP}.  The resummation has a simple exponential form
after transformation to the impact parameter, $b$, which is the
Fourier conjugate of $\qt $ \cite{PP}.  The necessary coefficients for
the inclusion of higher order terms in the resummation were calculated
in Ref.~\cite{DS} using the results of Ref.~\cite{EMP}. The formalism
which we shall use was written down in Ref.~\cite{CSS} using
techniques developed earlier for back-to-back jets \cite{CS}.
Numerical results of resummed calculations have been reported in
Refs.~\cite{AEGM,DWS}.

In order to have a complete description of the transverse momentum
one must match the theoretical results at large and small $\qt$.
The matching of the vector boson cross section
including $O(\alpha_S)$ has been considered in Ref.~\cite{AEGM}.
The matching including the full $O(\alpha_S^2)$ calculation appropriate
at large $\qt$
is given in Ref.~\cite{AK}.
The papers in Ref.~\cite{BR} have combined fixed order calculations with
a parton shower approach.

One of the most interesting results of the resummation procedure is
that for large enough vector boson mass perturbation theory is
valid even at $\qt=0$.  In fact, using the saddle point method it is
found \cite{PP,CSS} that the Fourier transform integral to recover the
transverse momentum distribution is determined (at $\qt=0$) by an impact
parameter
of order
\begin{equation}
b_{\rm SP} = \frac{1}{\Lambda} \Bigg(\frac{\Lambda}{Q} \Bigg)^{\kappa}\ ,
\end{equation}
where $\kappa=16/(49-2 n_f)$. Choosing $Q=M_W=80.33~\GeV, \Lambda=250~\MeV$
and the number of active flavours
$n_f=5$ we find that the saddle point is at the position
\begin{equation}
b_{\rm SP} ^{-1} \approx 10 \Lambda\ .
\end{equation}
Thus for $W$ and $Z$ production the integral is dominated by values of
$b$ which are at the borderline between the perturbative and
non-perturbative region.  Detailed predictions therefore depend
both on the perturbative Sudakov form factor and on a
parametrization of the non-perturbative part of the form factor,
to be extracted
from data.  The effect of non-perturbative terms on the
vector boson $\qt$ distribution has been considered in
Refs.~\cite{DWS,LY,R}.

A treatment of the production of vector bosons including both the
resummation at small transverse momentum and the decay kinematics has
been given in \cite{BQY}.  In this paper we carry the analysis
further and correct minor mistakes in Ref.~\cite{BQY}.  A more
complete theoretical description could be obtained if the order
$\alpha_S^2$ were included at large $\qt$, ({\it i.e.}~by extending the
results of Arnold and Kauffman \cite{AK} to include the decay of the
vector boson).

In Section \ref{cs} we give a brief summary of the formalism which we use
for the theoretical description of vector boson cross production.
Section \ref{ics} presents
results for the $O(\alpha_S)$ $\qt$-integrated cross-section.
Section \ref{results} contains our numerical results.
Besides discussing $W$ and $Z$ production,
we also review the available information on the non-perturbative
parameters which are important at low $\qt$, and
point out several issues which have not been addressed previously  in
the literature.  Our $\qt$-dependent predictions are obtained and
verified using two independent programs based on the formalism of Section
\ref{cs}. For the
$\qt$-integrated distributions we have implemented a separate program
based on the results of Section \ref{ics}.
Our conclusions are presented in Section \ref{conc}, while
details of the formulae used in our numerical
programs are given in two appendices.

\section{Resummed cross section at hadron level}
\label{cs}

In the Collins-Soper frame\footnote{The Collins-Soper frame
(defined in the Appendix \ref{details}) is the
rest frame of the vector boson with a specific choice for
orientation.} \cite{CS2}
the general expression for the resummed differential
cross section at hadron level may be written in the form
\bea
 { d \sigma(AB \rar V(\rar {l {\bar l'}}) X )
\over dq^2_T \, dQ^2 \, dy \,  d\cos{\theta} \, d\phi}
 &=&
{1 \over 2^8 N \pi S} \,
{Q^2 \over (Q^2 - \MB^2)^2 + \MB^2 \GB^2} \nonumber \\
&\times &
\bigg[Y_r(\qt^2, Q^2, y, \theta)
 + Y_f(\qt^2, Q^2,y, \theta, \phi) \bigg]  \ .
\label{totalcs}
\eea
In the above, $N=3$ is the number of colors,
$\sqrt{S}$ is the total hadron-hadron center-of-mass energy, while
$\theta$ and $\phi$ refer to the lepton polar and azimuthal angles.
The functions $Y_r$ and $Y_f$ stand for the resummed and
finite\footnote{The
``finite'' part is integrable as $\qt^2\rar 0$ and contains no
distributions.} parts of the cross
section,
respectively. They are defined in the following subsections.

\subsection{Resummed part}

The resummed part of the cross section is given as the Fourier integral
over the impact parameter $b$,
\bea
Y_r(\qt^2, Q^2, y, \theta) &=&  {1\over 2\pi}
\int_{}^{} b d b  \, J_{0}(\qt  b)
\sum_{a,b}{}^\prime
 F^{NP}_{a b } (Q,b,\xa,\xb) \nonumber \\
&\times &
W_{ab} (Q,b_*,\theta)
f'_{a/A}(\xa,\mu(b_*))
f'_{b/B}(\xb,\mu(b_*))
\ .
\label{resum}
\eea
In this expression the function $F^{NP}$
represents the non-perturbative part of the form factor. Its specific
form, as well as the definition of the variable $b_*$ will be
described below.
The variables $\xa$ and $\xb$ are given in terms of the
vector boson mass $Q$ and rapidity $y$ as
\beq
\xa = \frac{Q}{\sqrt{S}} \exp{(y)}\ ,\
\xb = \frac{Q}{\sqrt{S}} \exp{(- y)}\ .
\eeq
Note that
\beq
\tau = \xa \xb = Q^2/S\ .
\eeq

The modified parton structure functions $f'$
are related to the $\overline{MS}$ structure functions $f$
by a convolution,
\beq
f'_{a/A} (\xa,\mu) = \sum_{c}
\int_{\xa}^{1} {d z \over z} \,
C_{ac}( {\xa \over z},\mu )
f_{c/A} ( z, \mu )\ ,
\eeq
where ($a,b \neq g$) \cite{DS}
\bea
C_{ab}(z,\mu) &=&
\delta_{ab} \Bigg\{\delta(1-z)
+\frac{\alpha_S(\mu)}{2 \pi} C_F  \Big[ 1-z
+(\frac{\pi^2}{2}-4) \delta(1-z)\Big] \Bigg\} \ , \\
C_{ag}(z,\mu) &=& \frac{\alpha_S(\mu)}{2 \pi} T_R
\Big[ 2 z( 1-z) \Big]\ .
\eea
In the above the colour factors are $C_F=4/3$ and $T_R=1/2$, while
the prime on the sum in Eq. (\ref{resum})
indicates that gluons are excluded from the summation.

The function $W$ from Eq. (\ref{resum}) can be written in terms of the
the Sudakov form factor $\cS$ as
\beq
W_{ab} (Q,b,\theta)  = \exp{[\cS(b,Q)]} H^{(0)}_{ab} (\theta)\ .
\label{w}
\eeq
The exact definition of the Sudakov form factor will be discussed
below. The function
$H^{(0)}$, which includes the angular dependence
of the lowest order cross section and coupling factors,
is defined in Appendix \ref{details}.

\subsubsection{Sudakov form factor, large and small $b$}

In the formalism of Ref.~\cite{CSS} the Sudakov form factor
$\cS(b,Q)$ is given by
\beq
\cS(b,Q) = -\int_{b_0^2/b^2}^{Q^2}
{d {\bar \mu}^2\over {\bar \mu}^2}
\left[ \ln\left({Q^2\over {\bar \mu}^2}\right)
A\big(\alpha_s({\bar \mu})\big) +
B\big(\alpha_s({\bar \mu})\big) \right] \ ,
\label{Sudaffcs}
\eeq
with $b_0=2 \exp(-\gamma_E)\approx 1.1229$.
The coefficients $A$ and $B$ are perturbation series in $\alpha_S$,
\beq
A(\alpha_S) = \sum^\infty_{j=1} \left(\frac{\alpha_S}{2 \pi} \right)^j A^{(j)}\
, \
B(\alpha_S) = \sum^\infty_{j=1} \left(\frac{\alpha_S}{2 \pi} \right)^j B^{(j)}\
,
\eeq
where the first two coefficients in the expansion
are known \cite{DS}:
\bea
A^{(1)}&=&2 C_F\ ,\nonumber \\
A^{(2)}&=&2 C_F
\Big( N (\frac{67}{18}-\frac{\pi^2}{6})-\frac{10}{9} T_R n_f \Big)
\ ,\nonumber \\
B^{(1)}&=&-3 C_F\ ,\nonumber \\
B^{(2)}&=& C_F^2 \Big(\pi^2-\frac{3}{4}-12 \zeta(3)\Big)
      +C_F N \Big( \frac{11}{9} \pi^2-\frac{193}{12}+6 \zeta(3)\Big)
\nonumber \\
&+&C_F T_R n_f \Big(\frac{17}{3}-\frac{4}{9} \pi^2\Big)  \ .
\eea
In the above we take $n_f$ to be the number of quark flavors active at the
the scale at which $\alpha_S$ is evaluated.
In addition, the formalism of
Ref.~\cite{CSS} requires that the scale at
which the parton distributions are evaluated in
Eq.~(\ref{resum}) is
\beq
\mu(b)=b_0/b\ .
\label{cssscale}
\eeq

Equations (\ref{Sudaffcs}) through (\ref{cssscale}) should be compared
with the exact first order results \cite{AEGM} for the
Sudakov form factor,
\beq
\cS(b,Q) =\frac{\alpha_S}{2 \pi} C_F
\int_{0}^{Q^2}
{d {\bar \mu}^2\over {\bar \mu}^2}
\left[ 2 \ln\left({Q^2\over {\bar \mu}^2}\right) -3  \right]
\left[J_0(b  {\bar \mu}) -1 \right] \ ,
\label{Sudaff}
\eeq
and for the scale $\mu(b)$,
\beq
\mu(b) = Q \exp{\Bigl\{ - \int_0^Q \frac{dx}{x} [1-J_0(bx)] \Bigr\} }\ .
\eeq

Formally, the integration over $b$ in Eq.~(\ref{resum})
is from 0 to $\infty$.
However, as $b$ approaches $1/\Lambda$, the coupling $\alpha_S$
becomes large and the perturbative calculation of the form factor $\cS$
is no longer reliable.
This region is effectively removed from the integral by
evaluating $W$ and the parton structure functions at
\beq
b_* = \frac{b}{\sqrt{1+(b/\blim)^2}}\ ,
\label{blimeqn}
\eeq
which never exceeds the cut-off value $\blim$.
The large $b$ part of the Sudakov form factor is provided by the
function $F^{NP}$ which parametrizes the non-perturbative effects
\cite{DS,CSS,DWS}. The specific form of $F^{NP}$, as well as our
results with particular choices for non-perturbative parameters, will
be discussed in Section \ref{results}.

The formalism of Ref.~\cite{CSS} leaves open the question of small $b$.
The small $b$-region does not contribute large logarithms,
but a correct treatment is important to recover the
total cross section after integration over $\qt$.
The lowest order expression of Eq.~(\ref{Sudaff}) had the property that
$\cS\rightarrow 0$ as $b \rightarrow 0$, which is lost in
Eq.~(\ref{Sudaffcs}).  In Ref.~\cite{Davies} it has been  suggested
that one make the replacement in Eq.~(\ref{Sudaffcs})
\beq
\big(\frac{b_0}{b}\big)^2
\rightarrow \big(\frac{b_0}{b}\big)^2
\frac{1}{[1+b_0^2/(b^2 Q^2)]}\ ,
\eeq
This ensures that the scale never exceeds $Q$ by adding
power suppressed terms of order $(\qt/Q)^2$.

However, we shall use a slightly more sophisticated treatment
which ensures that
the first-order fixed-coupling result is correctly reproduced.
We define scales $\lambda$ and $\mu$ by
\bea
\int_\lambda^Q \frac{dx}{x} \ln\big(\frac{Q}{x} \big)
&=& \int_0^Q \frac{dx}{x}
\ln\big(\frac{Q}{x} \big) [1-J_0 (bx)]\ , \\
\int_\mu^Q \frac{dx}{x} &=& \int_0^Q \frac{dx}{x} [1-J_0(bx)]\ ,
\eea
which results in
\bea
\mu(b)&=&Q \exp{  \Big\{-\int_0^Q \frac{dx}{x} [1- J_0 (b x)] \Big\} } \ ,  \\
\lambda(b)&=&Q \exp{ \Big\{ - \Big[ 2 \int_0^Q \frac{dx}{x}
\ln \big(\frac{Q}{x} \big )
[1-J_0 (b x)] \Big]^{\frac{1}{2}} \Big\} } \ .
\eea

\begin{figure}[p]
\vspace*{+7.cm}
\includegraphics{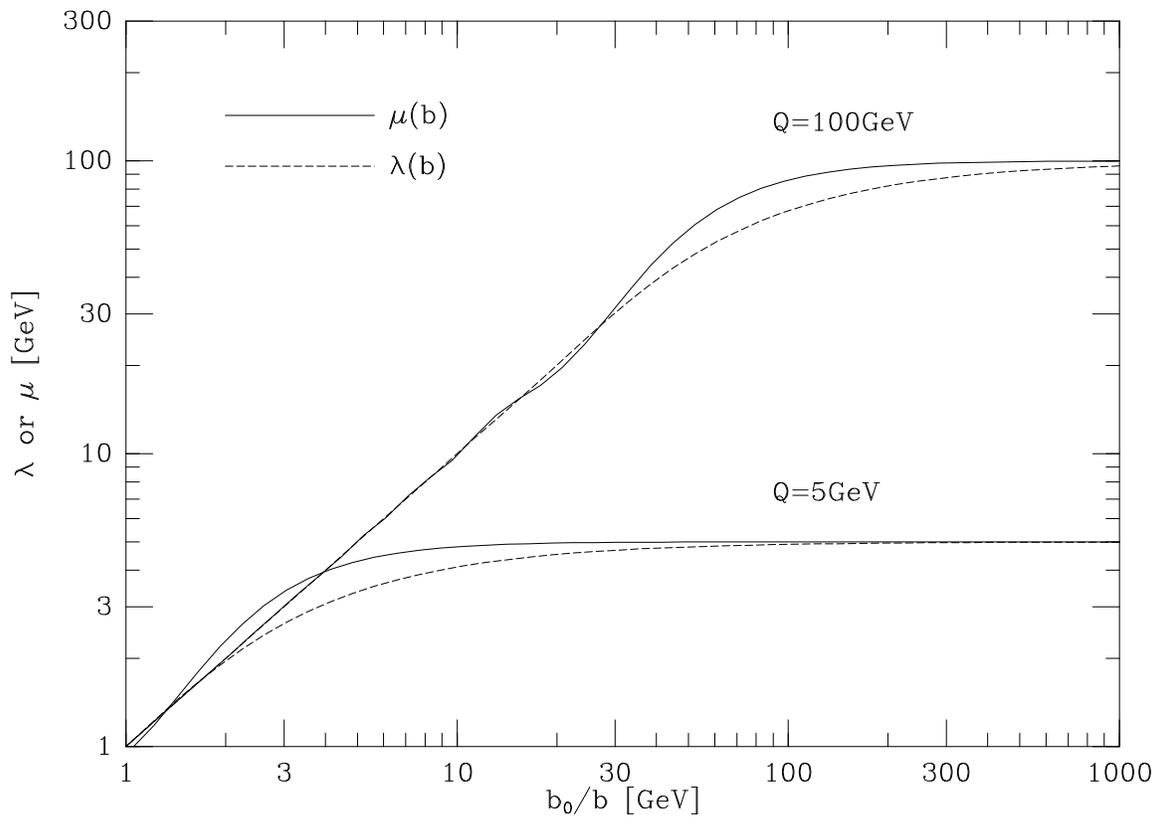}
\caption{Scales $\mu(b)$ and $\lambda(b)$ compared to $b_0/b$ for
$Q=5,100~\GeV$.}
\label{bscal}
\end{figure}

Figure \ref{bscal} shows  $\lambda$ and $\mu$ plotted for $Q=5$
and $100~\GeV$. At large $b \gg 1/Q$
where the resummation is mandatory, we have that
\beq
\lambda,\mu \sim \frac{b_0}{b}\ ,
\eeq
in accordance with Eq.~(\ref{Sudaffcs}) and the procedure of Ref.~\cite{CSS}.
In addition we have that $\mu(b),\lambda(b) \leq Q$,
where the equality is true for $b=0$.

\begin{figure}[p]
\vspace{12.0cm}
\includegraphics{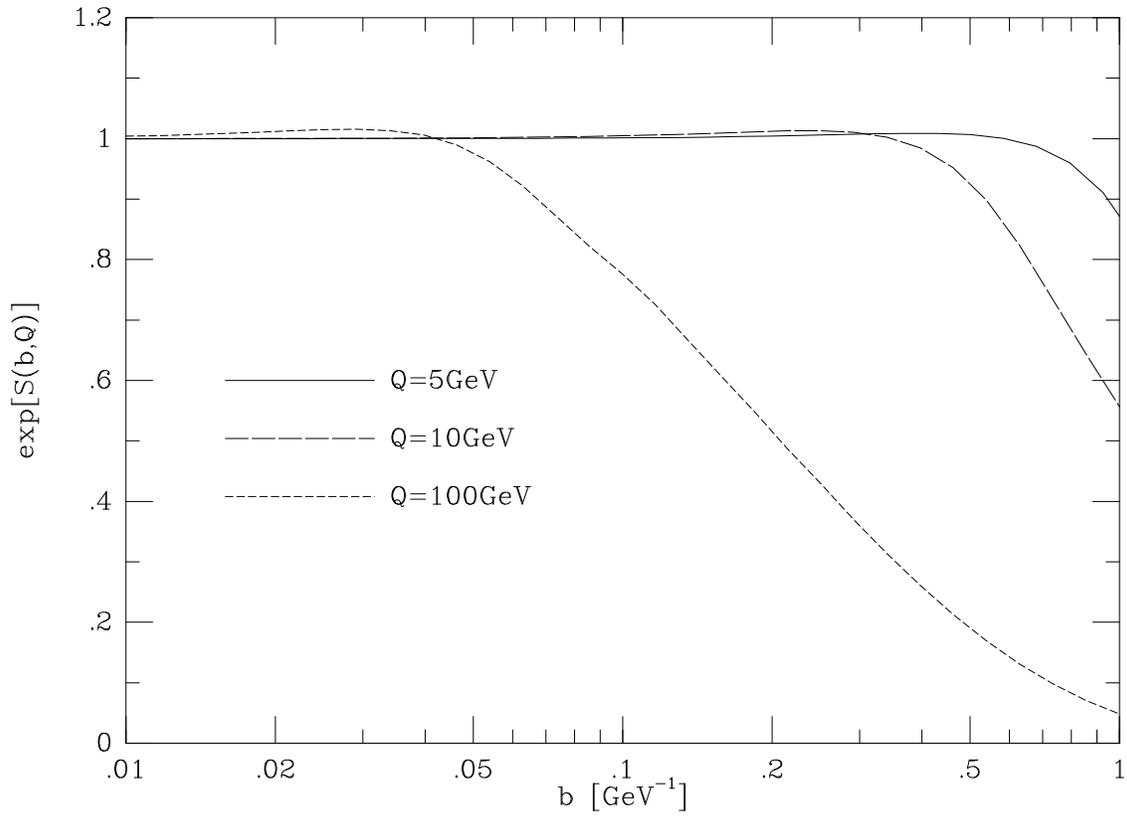}
\caption{The Sudakov form factor for $Q=5,10$ and $100~\GeV$.}
\label{sud}
\end{figure}

Therefore, instead of Eq. (\ref{Sudaffcs})
for the resummed form factor we shall use
\beq
\cS(b,Q) =
-\Bigg[\int_{\lambda^2(b)}^{Q^2}
{d {\bar \mu}^2\over {\bar \mu}^2}
\ln\left({Q^2\over {\bar \mu}^2}\right)
A\big(\alpha_s({\bar \mu})\big)
+\int_{\mu^2(b)}^{Q^2}
{d {\bar \mu}^2\over {\bar \mu}^2} B\big(\alpha_s({\bar \mu})\big)\Bigg]\ .
\label{Sudaffsplit}
\eeq
For fixed coupling constant this expression is exactly in agreement with the
lowest order result of Eq.~(\ref{Sudaff}), but also
preserves the good features of
Eq.~(\ref{Sudaffcs}).
The exponential of the Sudakov form factor with this prescription
is shown in  Figure~\ref{sud} for $Q=5,10$ and $100~\GeV$.

\subsection{Finite part}

The finite part of the cross section Eq.~(\ref{totalcs}) is defined as
\bea
Y_f(\qt^2,Q^2,y,\theta,\phi) &=&
\frac{1}{\pi Q^2}\int_{\xa}^{1} {d \za  \over \za }
\int_{\xb}^{1} {d \zb  \over \zb } \sum_{a,b}
\sum_{j=1}^{\infty} \left[{\alpha_s(Q) \over 2 \pi} \right]^j
\nonumber \\
&\times& R_{ab}^{(j)} (Q^2 ,\za,\zb,\theta,\phi)
\, f_{a/A}({\xa\over\za}  ,Q)  f_{b/B}({\xb\over\zb} ,Q)\  .
\label{finite}
\eea
To order $\cO(\alpha_S)$ we only need the function $R^{(1)}$, which
is given by the difference of the parts
derived from one parton emission,
and the pieces which have been removed from the
cross section in the factorization or resummation
procedure ($H^{(1)}$ and $\Phi^{(1)}$, respectively).
Therefore, we can write
\bea
R^{(1)}_{ab}(Q^2,\za,\zb,\theta,\phi)
 &=& Q^2 H^{(1)}_{ab}(Q^2,\za,\zb,\theta,\phi)\  \delta(s+t+u-Q^2)
\nonumber \\
&-&\Phi^{(1)}_{ab}(Q^2,\za,\zb,\theta)\ .
\eea
$H^{(1)}$ can further be separated into parts which are divergent
($S^{(1)}$) or
integrable as $\qt^2\rar 0$ ($\ovl{H}^{(1)}$), i.e.
\beq
H^{(1)}_{ab} =\ovl{H}^{(1)}_{ab}+S^{(1)}_{ab}\ .
\label{finite1}
\eeq
$S^{(1)}$ can then be combined with $\Phi^{(1)}$
into a function $\Sigma^{(1)}$,
so that the whole residue $R^{(1)}$, now in the form
\bea
R^{(1)}_{ab}(Q^2,\za,\zb,\theta,\phi)
&=&  Q^2 \ovl{H}^{(1)}_{ab}(Q^2,\za,\zb,\theta,\phi) \ \delta(s+t+u-Q^2)
\nonumber \\
& +&\Sigma^{(1)}_{ab}(Q^2,\za,\zb,\theta)\ ,
\label{finite2}
\eea
has the property
of being integrable as $\qt^2$ goes to zero. Explicit expressions
for functions $\ovl{H}^{(1)}$ and $\Sigma^{(1)}$, as well
as the relation of $Q^2$, $s$, $t$ and $u$
to $\za$ and $\zb$, are given in Appendix \ref{details}.

\section{Integrated cross section}
\label{ics}

With our definitions, after integration
over $\qt,\cos \theta$ and $\phi$, and dropping $\cO(\alpha_S^2)$ terms,
we recover exactly the order
$\cO(\alpha_S)$ cross section, which can be written in the form
\bea
\hspace*{-5mm} { d \sigma(AB \rar V X )
\over dQ^2 \, dy}
&\hspace*{-1mm} = \hspace*{-1mm}&
{1 \over 2^8 N \pi S} \,
{Q^2 \over (Q^2 - \MB^2)^2 + \MB^2 \GB^2} \frac{16}{3} \Bigg[
\sum_{a,b}{}^\prime
 f'_{a/A}(\xa,Q) f'_{b/B}(\xa,Q) \ \ \ \nonumber \\
&\hspace*{-1mm} + \hspace*{-1mm}& {\alpha_s(Q) \over 2 \pi}
\int_{\xa}^{1} {{d \za\over \za}   }
\int_{\xb}^{1} {{d \zb \over \zb} }
\sum_{a,b} X_{ab}(\za,\zb)  f_{a/A}({\xa \over \za} ,Q)
f_{b/B}({\xb \over \zb} ,Q)\Bigg]\ .
\label{rapcs}
\eea
In this case the modified parton distribution functions $f'$
are defined as ($a,b \neq g$)
\beq
f'_{a/A} (\xa,\mu) = \sum_{c}
\int_{\xa}^{1} {d z \over z} \,
D_{ac}( {\xa \over z},\mu )
f_{c/A} ( z, \mu )\ ,
\eeq
with coefficients
\bea
D_{ab}(z,\mu) &=&
\delta_{ab} \Bigg\{\delta(1-z)
+\frac{\alpha_S(\mu)}{2 \pi} C_F  \Bigg[
 1-z - \frac{1+z^2 }{1-z } \ln{ \Big( \frac{1+z}{2} \Big)  }
\nonumber \\
&+& (1+z^2) \Bigg( \frac{\ln (1-z)} {1-z}\Bigg)_+
+ (\frac{\pi^2}{2}-4) \ \delta(1-z) \Bigg] \Bigg\}\ , \\
D_{ag}(z,\mu) &=& \frac{\alpha_S(\mu)}{2 \pi}
T_R\Bigg[2 z (1-z) + [z^2 +(1-z)^2]
\ln \Big(\frac{2 (1-z)}{1+z} \Big) \Bigg] \ .
\eea
As before, the prime on the sum  indicates that gluons are excluded
from the summation.
The functions $X_{a b}$ are given as ($X_{\bq q}=
X_{q \bq}, X_{\bq g} = X_{q g}, X_{g \bq} = X_{g q}$)
\bea
X_{q \bq}(\za,\zb)&=&
G_{q\bq}(\za,\zb)\left( \frac{1}{(1-\za)(1-\zb)}\right)_{++}
+F_{q\bq}(\za,\zb) \ ,\\
X_{q g}(\za,\zb)&=&
G_{qg} (\za,\zb)\left(\frac{1}{1-\za}\right)_+
\ ,\\
X_{g q}(\za,\zb)&=&
G_{qg} (\zb,\za)\left(\frac{1}{1-\zb}\right)_+ \ ,
\eea
where we defined
\bea
F_{q\bq}(\za,\zb) &=&
  C_F\frac{  2 (1+\za \zb) (\za^2+\zb^2)}
{(\za+\zb)^2} \ ,\\
G_{q\bq}(\za,\zb) &=&
C_F \frac{2 (1+\za \zb)(\za^2 +\zb^2)}
 { (1+\za) (1+\zb)}
 \ ,\\
G_{q g}(\za,\zb) &=& T_R \frac{2 \za(1+\za \zb)} { (1+\za) (\za+\zb)}
\left[\frac{(1+\za \zb)^2 \zb^2}{(\za+\zb)^2} +(1-\za \zb)^2 \right]
 \ .
\eea
The definitions of the single and double ``plus''
distributions used in the above are given in Appendix \ref{details}.
We also remind the reader that  results in this section
are expressed in terms of $\overline{MS}$ scheme
structure functions. Similar expressions in the DIS scheme are presented in
Ref.~\cite{Kubar}.\footnote{We believe that the $qg$ contribution in
Ref.~\cite{Kubar}
is in error.}
Using the techniques of Ref.~\cite{Kubar} we can further integrate this to
the standard result for the total cross section \cite{ESW} in the
$\overline{MS}$ scheme.

\section{Results}
\label{results}

As already noted in the introduction, the motivation for
this work is  that a precise measurement of the $W$ mass
at a hadronic collider will depend on accurate theoretical information
about its production properties. However, before presenting our
results for $W$ and $Z$ production, we would like to address
the following issues which in our opinion have not been adequately
discussed in the literature:
the determination of the form of the non-perturbative
function from the low-energy Drell-Yan data, the dependence
of the results on the choice of $\blim$ and the matching between
low and high $\qt$.

\subsection{Determination of $F^{NP}$}

The  unknown function $F^{NP}$ from Eq.~(\ref{resum})
has a general form \cite{CSS}
\beq
F^{NP}_{ij} (Q,b,\xa,\xb) = \exp{
\left\{-\left[h_Q (b) \ln{\Big(\frac{Q}{2Q_0}\Big)}
+h_i (b,\xa) +h_j(b,\xb)\right] \right\} }\ ,
\label{FNP}
\eeq
where the functions $h$ are not calculable in perturbation
theory and therefore must be extracted from experiment.
On general grounds,  we expect
that $h_i \rar 0$  as $b \rar 0$, so that
the $\qt$-integrated  cross section is unchanged.
On the other hand, the parameter $Q_0$ is completely arbitrary.

The first attempt
to obtain $F^{NP}$ from experiment was made by
Davies et al. (DWS) in Ref. \cite{DWS}.  There
the functions $h$ were approximated by
\bea
&h_Q (b) = g_2 b^2\ ,&\nonumber \\
&h_i (b,\xa) +h_j(b,\xb)  =  g_1 b^2\ ,&
\label{dws_h}
\eea
since the observed
$\qt$ distribution at low $Q$ was approximately
gaussian in shape. Using the Duke and Owens parton distribution
functions \cite{DO},
DWS determined
parameters $g_1$ and $g_2$ from E288 \cite{E288} data with
$\sqrt{S}=27.4 \GeV$,  and also from R209 \cite{R209}
data. The resulting values (with a particular choice of $Q_0 = 2\GeV$) were
\beq
g_1 = 0.15\GeV^2\ ,\ g_2 = 0.40\GeV^2 \ .
\label{dws_g}
\eeq
The cut-off value of $\blim $ from Eq.~(\ref{blimeqn}) was
chosen to be $0.5\GeV^{-1}$.
This parameter set yielded a good agreement of theory
with R209, as well as with E288 data for $ Q< 9 \GeV$ mass bins.
However, for $Q > 11\GeV$ theoretical expectations were unacceptably
far above the data.

Motivated by the fact  that the production of vector
bosons at Fermilab Tevatron ($\sqrt{S}=1.8\TeV$) involves
values of $\tau = \xa \xb $ which are significantly
lower than those considered in Ref. \cite{DWS}, Ladinsky and Yuan (LY)
\cite{LY} reinvestigated the form of the non-perturbative functions from
Eq.~(\ref{FNP}). Using  CTEQ2M parton distribution
functions \cite{CTEQ2},  they showed that the DWS form of non-perturbative
function (Eqs. (\ref{dws_h} and (\ref{dws_g}))
no longer agrees with R209 data for $5 \GeV < Q < 8 \GeV$.
In order to improve theoretical predictions, LY postulated
the $\tau$ dependence for functions $h_i$, so that
\bea
h_Q (b) &=& g_2 b^2\ ,\nonumber \\
h_i (b,\xa) +h_j(b,\xb)  &=&  g_1 b^2 +
g_1 g_3 b \;\ln\Big({\tau\over \tau_0}\Big)\ ,
\label{ly_h}
\eea
where $\tau_0$ is arbitrary parameter.\footnote{We note here that,
unlike a simple gaussian,
the particular functional
form of the non-perturbative function in Eq.~(\ref{ly_h})
is not always positive.} Choosing $\tau_0 = 0.01$,
$Q_0 = 1.6\GeV$,
and $\blim  = 0.5\GeV^{-1}$, these authors
determined the non-perturbative parameters
by comparison of theory
to R209 data \cite{R209} in the range $5\GeV < Q < 8\GeV$,
to $\sqrt{S}=27.4\GeV$ E288 data \cite{E288} in the range
$6\GeV < Q < 8\GeV$, and also to CDF $Z$ data \cite{CDFZ}.
In particular, LY explicitly showed that  values of
\beq
g_1=0.11^{+0.04}_{-0.03}\GeV^2\ ,\; g_2=0.58^{+0.1}_{-0.2}\GeV^2\ ,\;
g_3=-1.5^{+0.1}_{-0.1}\GeV^{-1}\ ,
\label{ly_g}
\eeq
provide a good agreement of theory with CDF $Z$ and with
R209 data (for $5\GeV < Q < 8\GeV$),
as well as with CDF $W$ data  \cite{CDFW}. They
furthermore noted that the parameters of Eq.~(\ref{ly_g})
give results which are in agreement with E288 data
and with R209 data for $11\GeV < Q < 25\GeV$.

The problem with using the Drell-Yan data from the
fixed-target experiments is that the overall
normalization of the cross section  is uncertain.
For example, the E288 data has a stated
normalization uncertainty of 25\% \cite{E288}.
Since smearing function $F^{NP}$
simply shifts the $\qt$ distribution between the low and high $\qt$
regions, and since the bulk of the data is in the low
$\qt$ region, it is clear that using data with  wrong normalization
will affect the non-perturbative parameters.
Therefore, it is necessary to establish correct normalization for
fixed-target experiments before trying to determine the shape of $F^{NP}$
from their data.

A consistent way of determining overall normalization
is to compare   theoretical predictions for
$\qt$-integrated cross section (to a given order in $\alpha_S$)
with experimental results. We illustrate that procedure
for E288 \cite{E288} and
E605 \cite{E605} experiments.\footnote{Unless otherwise
stated, all results described
in this section are obtained using CTEQ2M parton distribution
functions, which facilitates comparison to previous work \cite{LY}.}

\begin{figure}[p]
\vspace{12.0cm}
\includegraphics{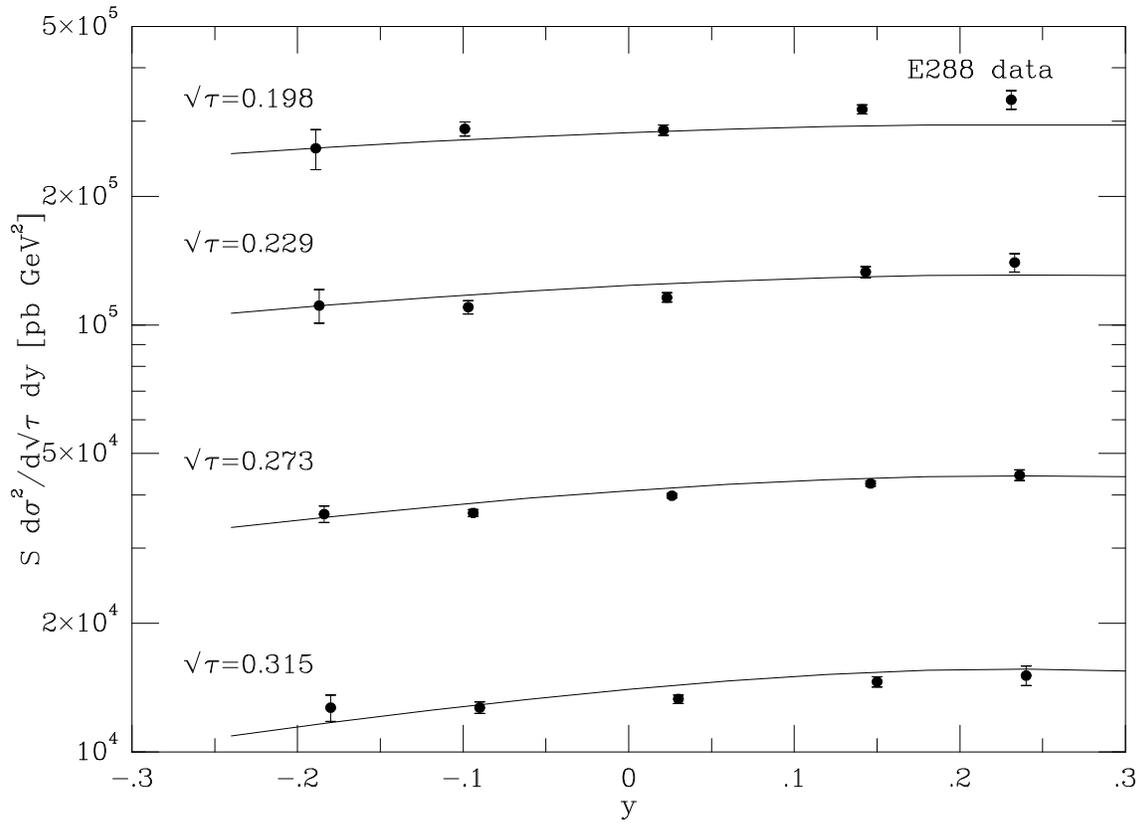}
\caption{$S\ d\sigma^2 /d\protect\sqrt{\tau}dy$
distribution from E288 compared to
theory multiplied by  $K=0.83$.}
\label{e288_y}
\end{figure}

\begin{figure}[p]
\vspace{12.0cm}
\includegraphics{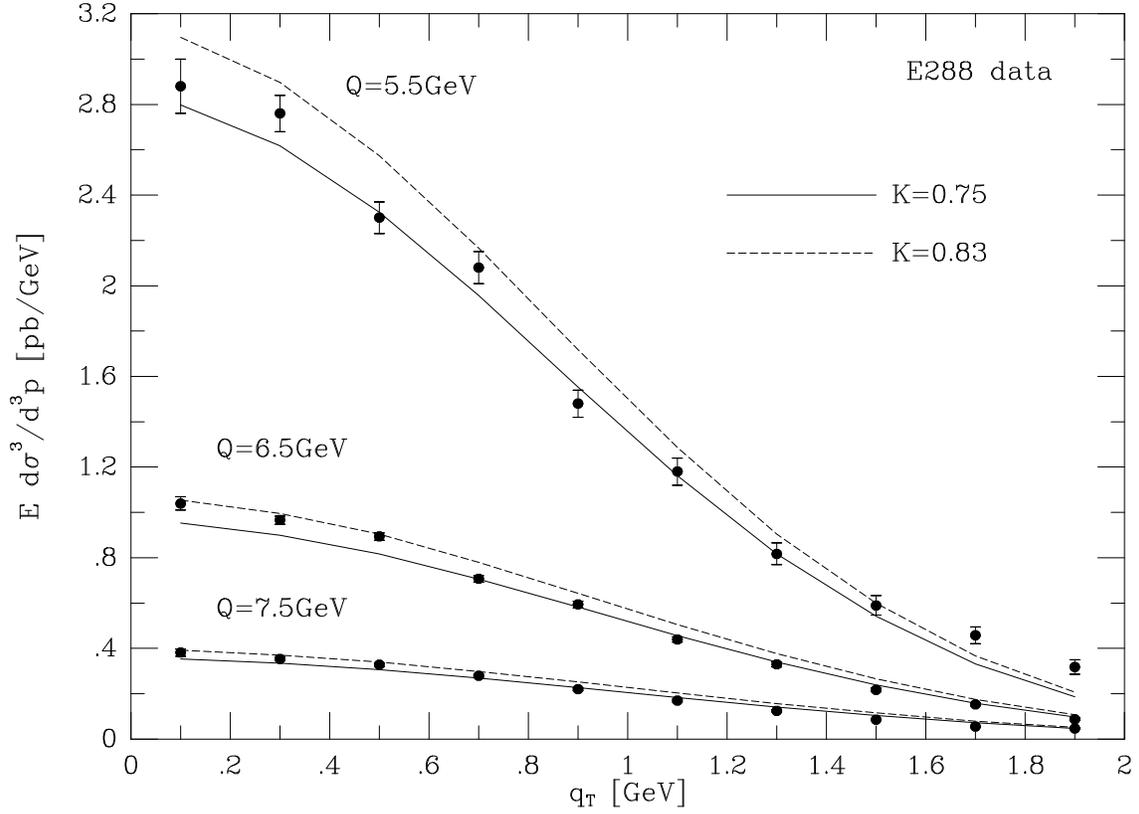}
\caption{$E\ d\sigma^3 /d^3p$
distribution from E288 compared to
theory with $K=0.75$
(full line) and $K=0.83 $ (dashed line).}
\label{e288_qt}
\end{figure}

For E288 ($\sqrt{S}=27.4 \GeV$) we used
$S\ d\sigma^2 /d\protect\sqrt{\tau}dy$ distributions
for the $\sqrt{\tau}$ bins with
$Q<9 \GeV$.\footnote{The $\sqrt{\tau}$ bins with $Q>11\GeV$ were discarded
because
of the low statistics.}
In order to achieve agreement with experiment (see Figure \ref{e288_y}),
we found that theoretical results had to be rescaled down by a factor
$K= 0.83\pm 0.03$.\footnote{We define the $K$-factor as
$K=experiment/theory$. We have chosen to change the normalization
of the theory, rather than shifting the experimental data, despite the fact
that it is the data which is subject to a normalization uncertainty.}
On the other hand, for the invariant cross section versus $\qt$
data in the range $5\GeV < Q < 9\GeV$ and $\qt < 2\GeV$,
the central values of LY parameters
yield the best $\chi^2$ of about $7.7/dof$
for $K=0.75$, while $K$-factors of 0.80 and 0.83 lead to $\chi^2$
of about $11.9/dof$ and $18.2/dof$, respectively.
These results, shown in Figure \ref{e288_qt},
indicate that parameters of Eq.~(\ref{ly_g})
overestimate the E288 transverse momentum data.
We are therefore unconvinced that the parameters
of Eq.~(\ref{ly_g}) give the best possible fit to the data.

\begin{figure}[p]
\vspace{12.0cm}
\includegraphics{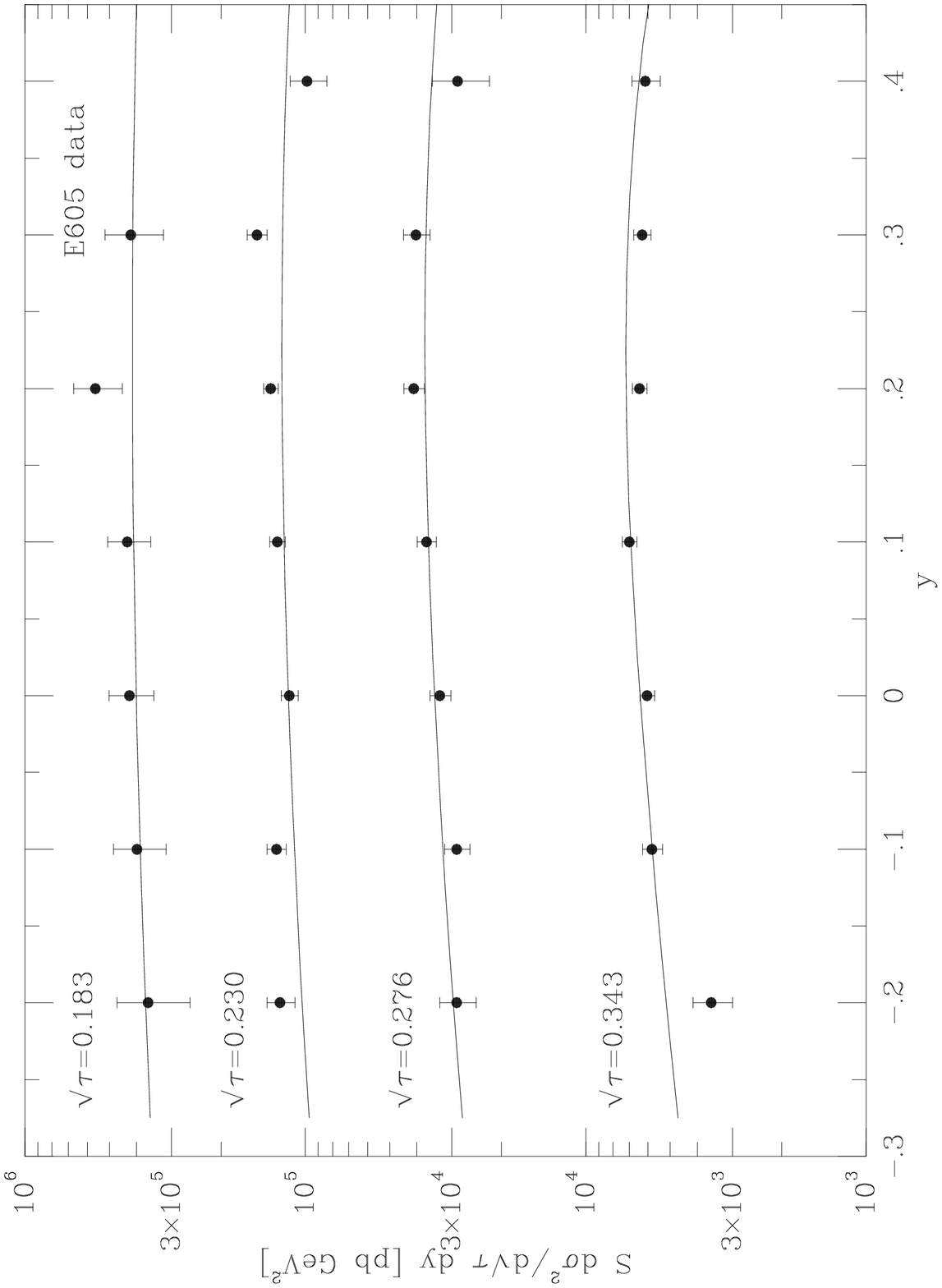}
\caption{$S\ d\sigma^2 /d\protect\sqrt{\tau}dy$
distribution from E605 compared to
theory multiplied by  $K=0.88$.}
\label{e605_y}
\end{figure}

\begin{figure}[p]
\vspace{12.0cm}
\includegraphics{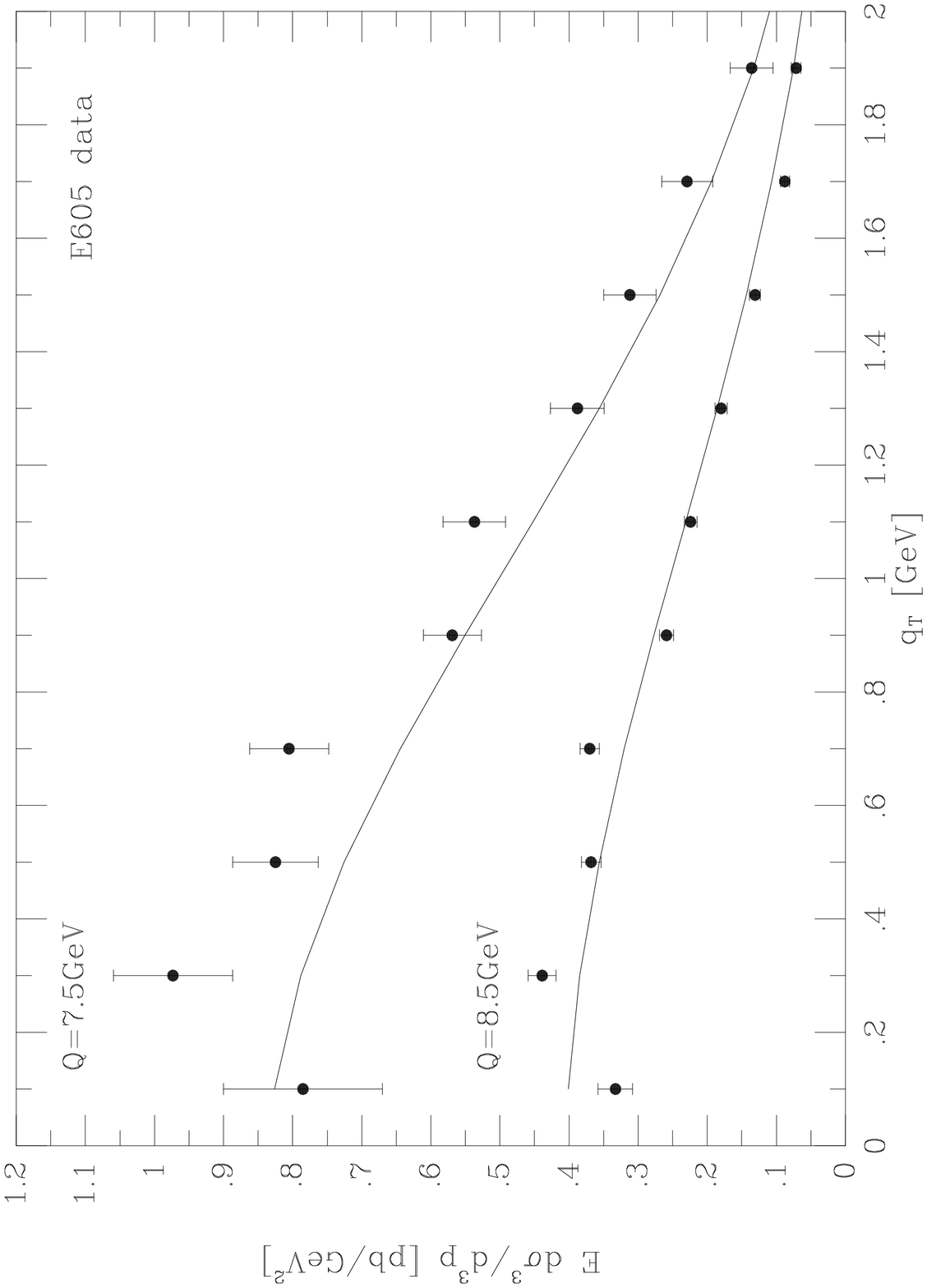}
\caption{$E\ d\sigma^3 /d^3p$
distribution from E605 (mass bins with $Q<9\GeV$) compared to
theory with $K=0.88$.}
\label{e605_qt}
\end{figure}

\begin{figure}[p]
\vspace{12.0cm}
\includegraphics{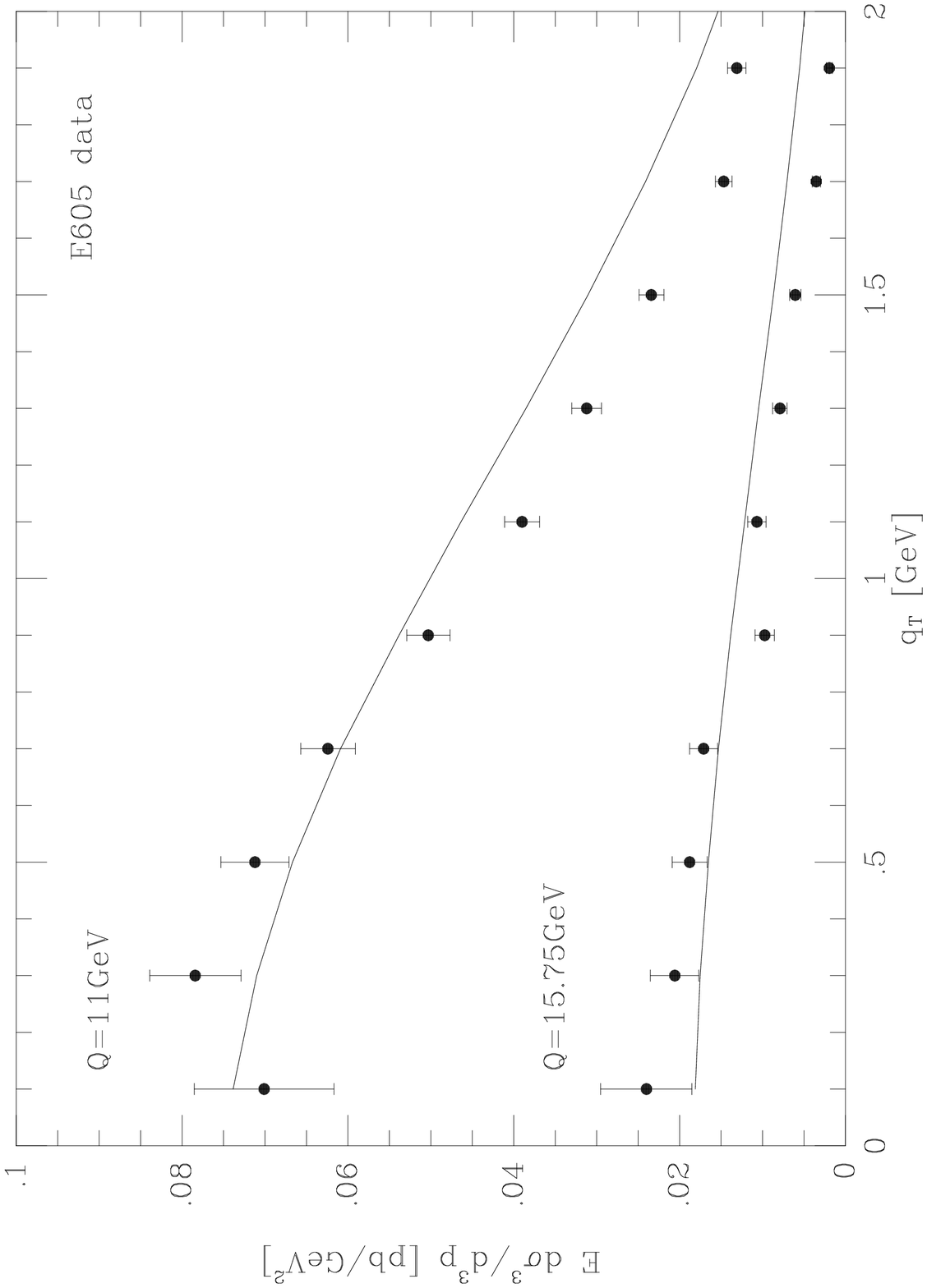}
\caption{$E\ d\sigma^3 /d^3p$
distribution from E605 (mass bins with $Q>10.5\GeV$) compared to
theory with $K=0.88$.}
\label{e605_qt_2}
\end{figure}

For E605 ($\sqrt{S}=38.7 \GeV$) we used all available
$\sqrt{\tau}$ bins for  $S\ d\sigma^2 /d\protect\sqrt{\tau}dy$ distributions,
and determined $K$-factor of $K= 0.88\pm 0.02$.\footnote{E605 experiment
has a stated normalization uncertainty of 15\%.}
The agreement between the theory and the data is illustrated in
Figure \ref{e605_y}. The corresponding invariant cross section
distributions obtained with central values of
LY parameters are shown in Figures
\ref{e605_qt} and \ref{e605_qt_2}. From Figure \ref{e605_qt} it can be seen
that  theoretical predictions are in good agreement
with experimental results for the bins with $Q<9\GeV$. However,
for bins with $Q>10.5\GeV$ (Figure \ref{e605_qt_2})
the theoretical distributions do not match the distributions
obtained by experiment.

The LY functional form of $F^{NP}$ given in Eq. (\ref{ly_h}) implies
that theory should be able to describe data with different values
of $\tau$. Our results certainly do not support that statement, and we
believe that the form of the non-perturbative function remains
an open question.

\subsection{Choice of $\blim $}

Another issue which has not been addressed in the literature
is the dependence of results on the
choice of the cut-off value $b_{\lim}$ from
Eq. (\ref{blimeqn}). In the original work of Ref. \cite{DWS} $\blim $
was taken to be $0.5\GeV^{-1}$ because the structure functions
were not defined for scales less than $2\GeV$.
However, this choice is arbitrary.
Any change in $\blim $ (in a reasonable range around $0.5\GeV^{-1}$)
should be compensated by a change in
non-perturbative parameters describing $F^{NP}$, so that
an equally good description of experimental data is always achieved.

\begin{figure}[p]
\vspace{12.0cm}
\includegraphics{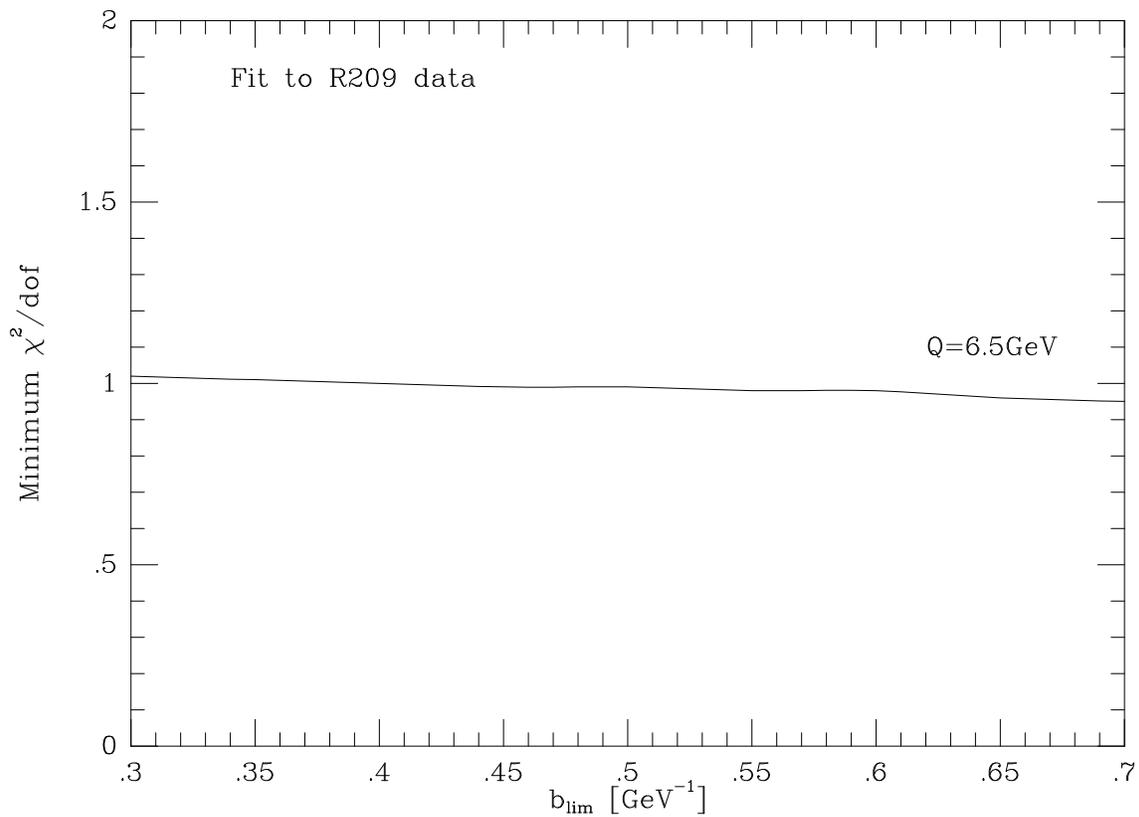}
\caption{Best $\chi^2/dof$ obtained by varying
$ g$ for R209 data ($5\GeV<Q<8\GeV$ and
$\protect\qt<3\GeV$). For comparison, with
the LY non-perturbative function we obtained $\chi^2/dof$ of about $1.3$.
}
\label{blim_r209}
\end{figure}

\begin{figure}[p]
\vspace{12.0cm}
\includegraphics{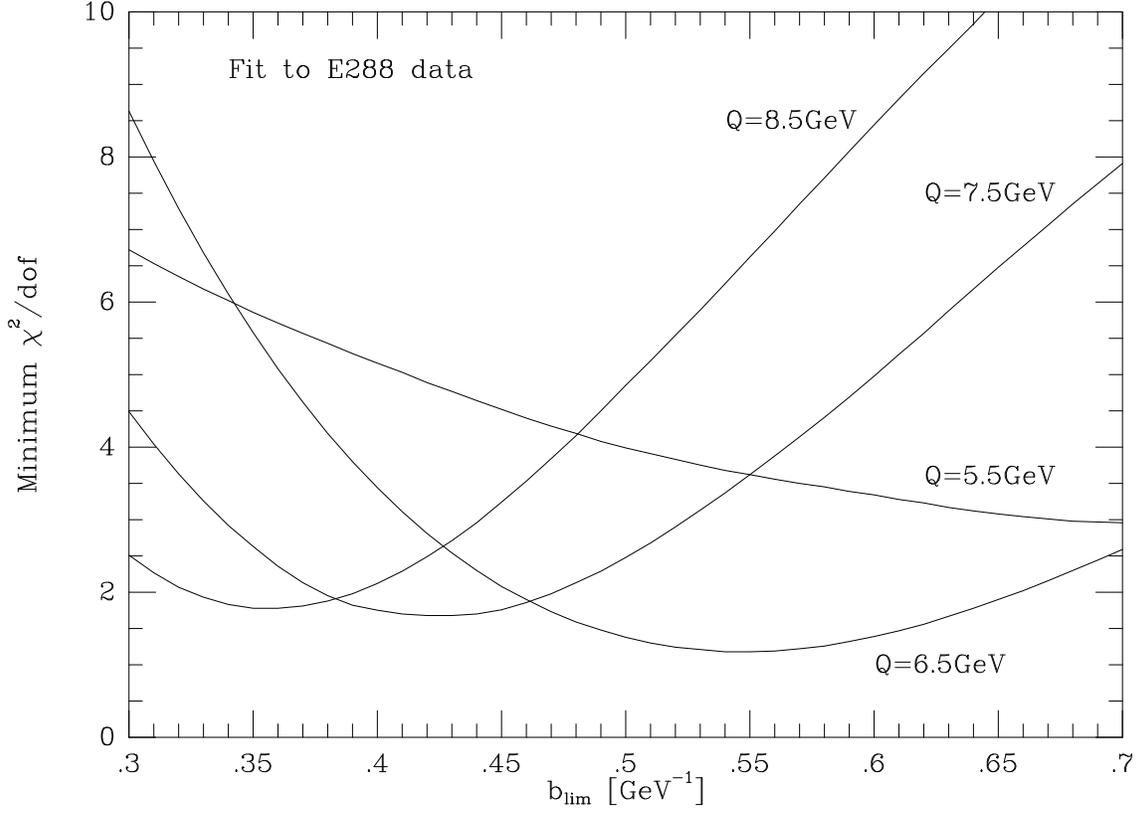}
\caption{Best $\chi^2/dof$ obtained by varying $ g$ for
E288 data. We used four data sets with $\protect\qt<2\GeV$
and $Q$ below $9\GeV$: $5\GeV<Q<6\GeV$, $6\GeV<Q<7\GeV$,
$7\GeV<Q<8\GeV$, and $8\GeV<Q<9\GeV$.
Theoretical results were multiplied by $K=0.83$.
For these data sets
the LY form of the non-perturbative function
yields $\chi^2/dof$ of 6.5, 14.1, 22.9 and 27.5, respectively.
}
\label{blim_e288}
\end{figure}

\begin{figure}[p]
\vspace{12.0cm}
\includegraphics{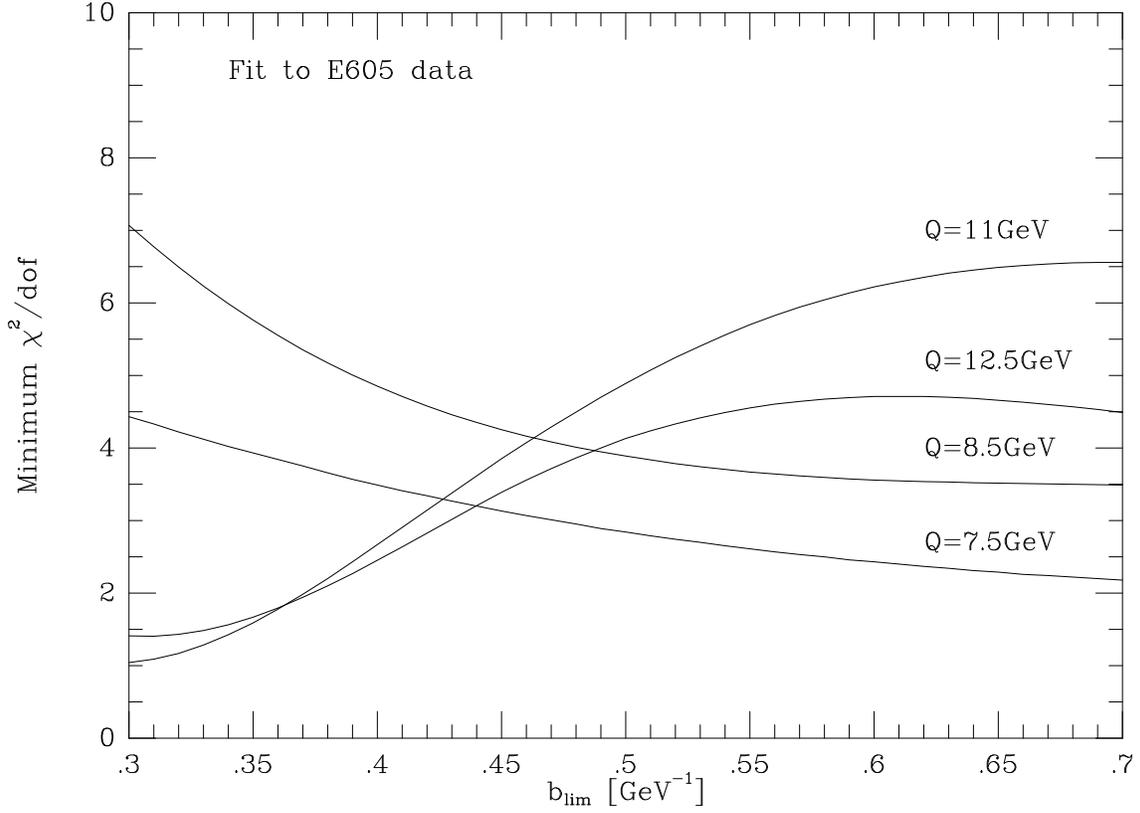}
\caption{Best $\chi^2/dof$ obtained by varying
$ g$ for
E605 data. We used data
sets for $7\GeV<Q<8\GeV$, $8\GeV<Q<9\GeV$,
$10.5\GeV<Q<11.5\GeV$ and $11.5\GeV<Q<13.5\GeV$
(with $\protect\qt<2\GeV$ and $K=0.88$).
For these data sets the LY form of the non-perturbative function
yields $\chi^2/dof$ of 2.2, 3.5, 16.6 and 17.3, respectively.
}
\label{blim_e605}
\end{figure}

In order to verify that statement, we attempt
to reproduce several different sets of transverse momentum distributions,
with  $\blim $ chosen
in the range from $0.3\GeV^{-1}$ to $0.7\GeV^{-1}$, which one
may consider to be reasonable.\footnote{Since CTEQ2M
structure functions are defined for scales greater than $1.6\GeV$,
and since $\mu(b)\simeq b_0/b$ for large $b$, the upper limit of
$0.7\GeV^{-1}$ is still a
possible choice for $\blim $.}
We have chosen $\qt$-distributions that include
R209 $d\sigma/d\qt^2$ data ($5\GeV < Q < 8\GeV$, with $\qt < 3\GeV$),
E288 $Ed\sigma/d^3p$ data
for $Q=5.5,6.5,7.5$ and $8.5\GeV$
(with $\qt < 2\GeV$ and $K$-factor of 0.83),
and also E605 $Ed\sigma/d^3p$ data
for $Q=7.5,8.5,11$ and $12.5\GeV$
(with $\qt < 2\GeV$ and $K$-factor of 0.88).
Since these data were obtained within a
narrow mass range we can neglect the $Q$-dependence of Eq. (\ref{FNP}),
and adopt a simple gaussian form for $F^{NP}$,
\beq
F^{NP} = \exp{(- g \ b^2)}\ ,
\label{gauss}
\eeq
where $ g $ is an effective parameter, different for each
of the above $\qt$ distributions. In this way, the task of
finding the form of the non-perturbative function is
reduced to a simple one-parameter fit. In Figures \ref{blim_r209},
\ref{blim_e288} and \ref{blim_e605}
we show the best $\chi^{2}/dof$ obtained by varying $ g $
with different values of $\blim $,
for R209, E288 and E605 data sets, respectively.
It is obvious that for
R209 data (where $\sqrt{\tau}=0.105$) an equally good fit
can be obtained regardless of the value chosen for $\blim $,
in part because there is only a small number of data
points of limited statistical precision.

However,
for E288 and E605 data, which involve larger
values of $\sqrt{\tau}$
(for E288 $\sqrt{\tau}=0.201, 0.238, 0.274$ and $0.311$, while for
E605  $\sqrt{\tau}= 0.194, 0.220, 0.284$ and $0.323$),
the choice of $\blim $ makes a considerable difference, which
is not satisfactory from the theoretical point of view.
These results might indicate that perhaps a pure gaussian
form of the non-perturbative function is wrong.\footnote{There is
no reason in principle
why the functions
$h$ from Eq. (\ref{FNP}) should be limited to a
quadratic dependence on $b$.} In any case, it is clear
that this problem requires a further study.

In Table \ref{g_range} we also note the range of values
for $g$ for which results
shown in Figures \ref{blim_r209} through \ref{blim_e605}
were obtained.
If the gaussian form of the non-perturbative function
were correct, then the values of $g$, which increase for increasing
$Q$ in the E288 and E605
sets,
would indicate that the effective coefficient in front
of  $\ln{(Q/(2 Q_0))}$ in Eq. (\ref{FNP}) should
be positive, and furthermore that the $\tau$ dependence
of $F^{NP}$ is not large.
However, the values of $g$ obtained
for R209 data do not seem to support that observation.

\subsection{Matching of low and high $\protect\qt$ regions}

The resummation formalism is expected to give a good
theoretical description of vector boson production
in the low $\qt$ region ($\qt^2 \ll Q^2$).
On the other hand, conventional perturbation theory
provides a good approximation in the other regime, for
$\qt^2 \gg Q^2$. The necessity of matching low and high $\qt$
regions has already been discussed in Ref. \cite{AK} for the
$W$ and $Z$ production, where it was shown that the
proper matching of the pure perturbative and resummed
expressions reduces theoretical errors.
However, matching will never be perfect
since resummation introduces higher order terms
which will not be cancelled at large $\qt$ in any finite-order
calculation of $Y_f$ from Eq. (\ref{totalcs}).  For this reason,
the conclusion of Ref. \cite{AK} was that one should prefer the ordinary
perturbation theory result once the resummed part $Y_r$
becomes negative. For the $W$ and $Z$ production
with $\sqrt{S}=1.8\TeV$ this happens at $\qt \sim 50\GeV$.
In Ref. \cite{AK} it was also shown that matching
works well with the second-order calculation of $Y_f$. For example, at
$\qt$ of about $50\GeV$ the mismatch between  the
$\qt$ distribution calculated using resummation plus the $\cO(\alpha_S^2)$
calculation of $Y_f$ and the one calculated using conventional
second-order perturbation theory was of the order of 10\%.

\begin{figure}[p]
\vspace{12.0cm}
\includegraphics{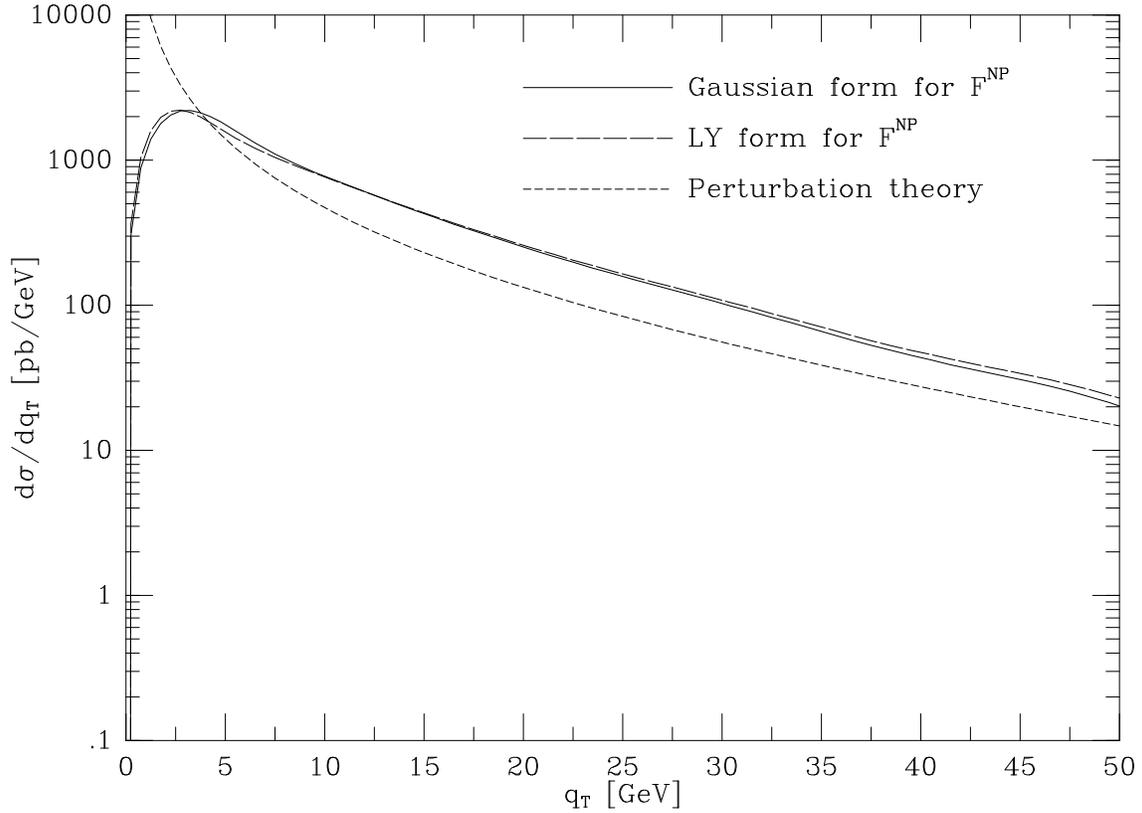}
\caption{Comparison of the theoretical $\protect\qt$ distributions
for $W=W^++W^-$ production at Tevatron ($\protect\sqrt{S}=1.8\protect\TeV$)
with $\cO(\alpha_S)$ perturbative calculation. These results
were obtained with  gaussian ($\protect\blim= 0.5\GeV^{-1}$ and
$g= 0.33\GeV^2$) and LY form of $F^{NP}$. We assumed
$BR(W\rar e\nu) = 0.111$.
}
\label{match_2}
\end{figure}

\begin{figure}[p]
\vspace{12.0cm}
\includegraphics{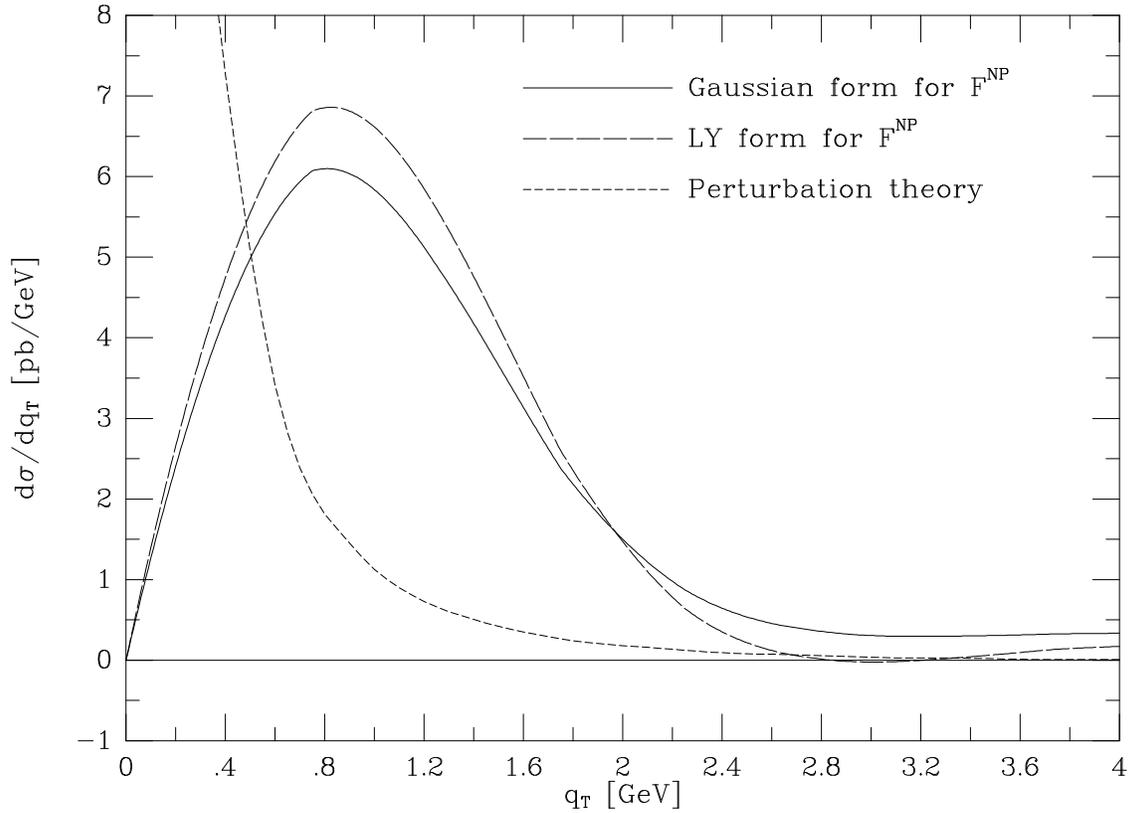}
\caption{Comparison of the theoretical (resummed plus finite)
$\protect\qt$ distributions
for E288 experiment ($\protect\sqrt{S}=27.4\protect\GeV$,
$5\GeV < Q < 6 \GeV$ and $-0.27 < y < 0.33$) with
$\cO(\alpha_S)$ perturbative calculation. These results were obtained with
gaussian ($\protect\blim= 0.5\GeV^{-1}$ and $g= 0.33\GeV^2$) and
LY form of $F^{NP}$.
}
\label{match}
\end{figure}

In Figure \ref{match_2} we show our results  for the $W$
production, obtained with gaussian and LY form of $F^{NP}$, compared
to the $\cO(\alpha_S)$ perturbative result. At $\qt$ of $50\GeV$
the mismatch between $Y_r$ plus the $\cO(\alpha_S)$ calculation of $Y_f$ and
perturbation theory is about $50\%$.
For the $\gamma^*$ production at fixed target experiments
the same problem (but even more acute) is illustrated in Figure \ref{match}
for E288 experiment (with $5\GeV < Q < 6\GeV$ and $-0.27<y<0.33$).
These two figures clearly indicate the necessity for
extending the results of Ref. \cite{AK} to include  the decay of the
vector boson, if a complete theoretical description is desired.
Nevertheless, we emphasize that results presented in this paper
are obtained for the low $\qt$ region, where the resummation
formalism plus  the $\cO(\alpha_S)$  calculation of $Y_f$ should be adequate.
Note also that it is the region of low $\qt$ which is of interest for
the $W$ mass measurement.

\subsection{Results for the $W$ and $Z$ production}

Once the high statistics data on the vector boson
$\qt$ distributions at $1.8 \TeV$ become available, one can
extract the effective form of $F^{NP}$  from the $Z$ data, and
use that information to obtain an accurate theoretical prediction for
the various $\qt$-dependent distributions of the $W$ boson.\footnote{Taking
current
values of vector boson masses \cite{PDG},
at $\sqrt{S}=1.8\TeV$ for $Q=M_W(M_Z)$
the value of $\sqrt{\tau}$ is about $0.045 (0.051)$.}
This in turn should allow a precise measurement of the $W$ mass.

\begin{figure}[p]
\vspace{12.0cm}
\includegraphics{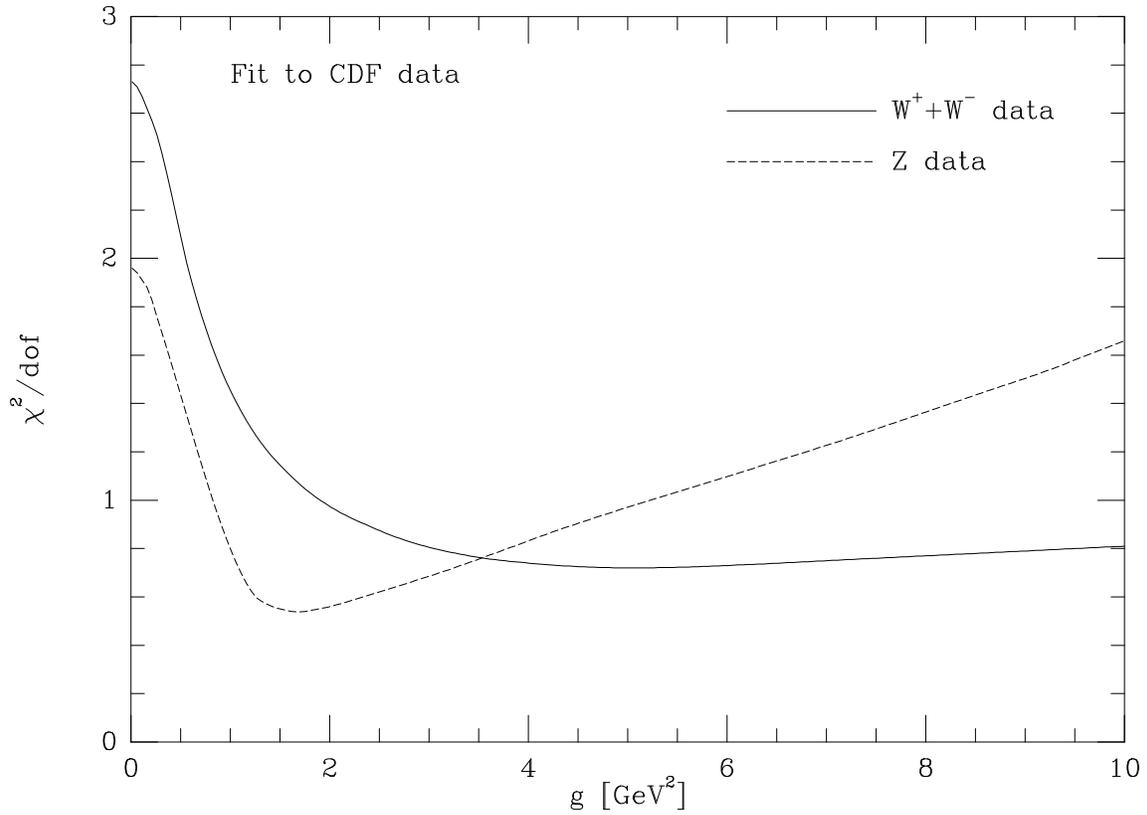}
\caption{$\chi^2/dof$ obtained by varying $g$ for CDF
$W^++W^-$ (full line) and $Z$ $d\sigma /d\protect \qt$
data (dashed line) below $\protect \qt = 45\GeV$. LY form of
the non-perturbative function yields $\chi^2/dof$ of about 0.6 and 0.5
for $W^++W^-$ and $Z$ data, respectively.
}
\label{chi2}
\end{figure}

For the data available at  present
\cite{CDFZ, CDFW} the statistics is low, which limits the predictive
power of the resummation formalism. In order to illustrate that, we again
adopt a simple parametrization of $F^{NP}$ given in Eq. (\ref{gauss}),
choose $\blim = 0.5\GeV^{-1}$,
and vary $g$ in an attempt to obtain a good fit to
the $W=W^++W^-$ and $Z$ $d\sigma /d\qt$ data.\footnote{We assumed
$BR(Z\rar e^+e^-) = 0.033$ and $BR(W^+\rar e^+\nu) = 0.111$, as
was done in \cite{CDFZ} and \cite{CDFW}, respectively.}
As shown in Figure \ref{chi2},
due to large statistical errors
almost any choice of $g$ in the range between of about
$1.5\GeV^2$ to about $5\GeV^2$
yields an acceptable description of both data sets. We
conclude that it does not make much sense to use
these data for determination
of $F^{NP}$, and that data with much higher statistics are
needed before any firm theoretical predictions can be made for
the $W$ and $Z$ production.
Nevertheless, we still observe
that the effective
value for $g$ obtained from the $Z$ data tends to be
smaller than the one obtained from the $W$ data. This is
in disaccord with what one would expect from the fixed target data
(see Table \ref{g_range}), and may again indicate that pure gaussian
form for $F^{NP}$ is not correct. It may also indicate experimental
biases introduced by the selection of two isolated leptons in the $Z$
sample.

\begin{figure}[p]
\vspace{12.0cm}
\includegraphics{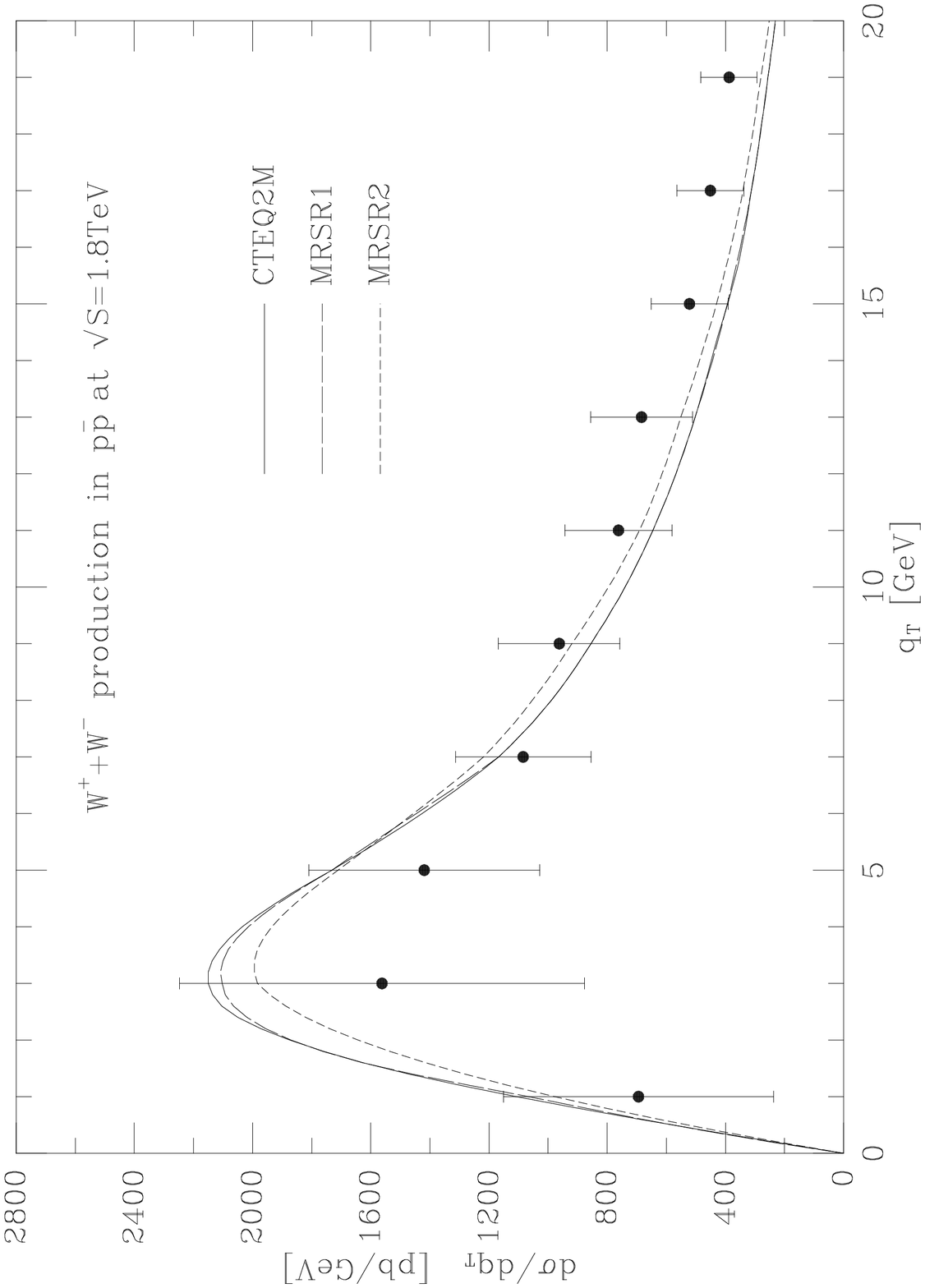}
\caption{Theoretical prediction  for CDF
$W^++W^-$ $d\sigma /d\protect \qt$ data. These results are
obtained using several different parton distribution functions, and with
an effective gaussian form of $F^{NP}$ ($g = 3.0\GeV^2,\blim=0.5\GeV^{-1}$).
}
\label{wpm}
\end{figure}

\begin{figure}[p]
\vspace{12.0cm}
\includegraphics{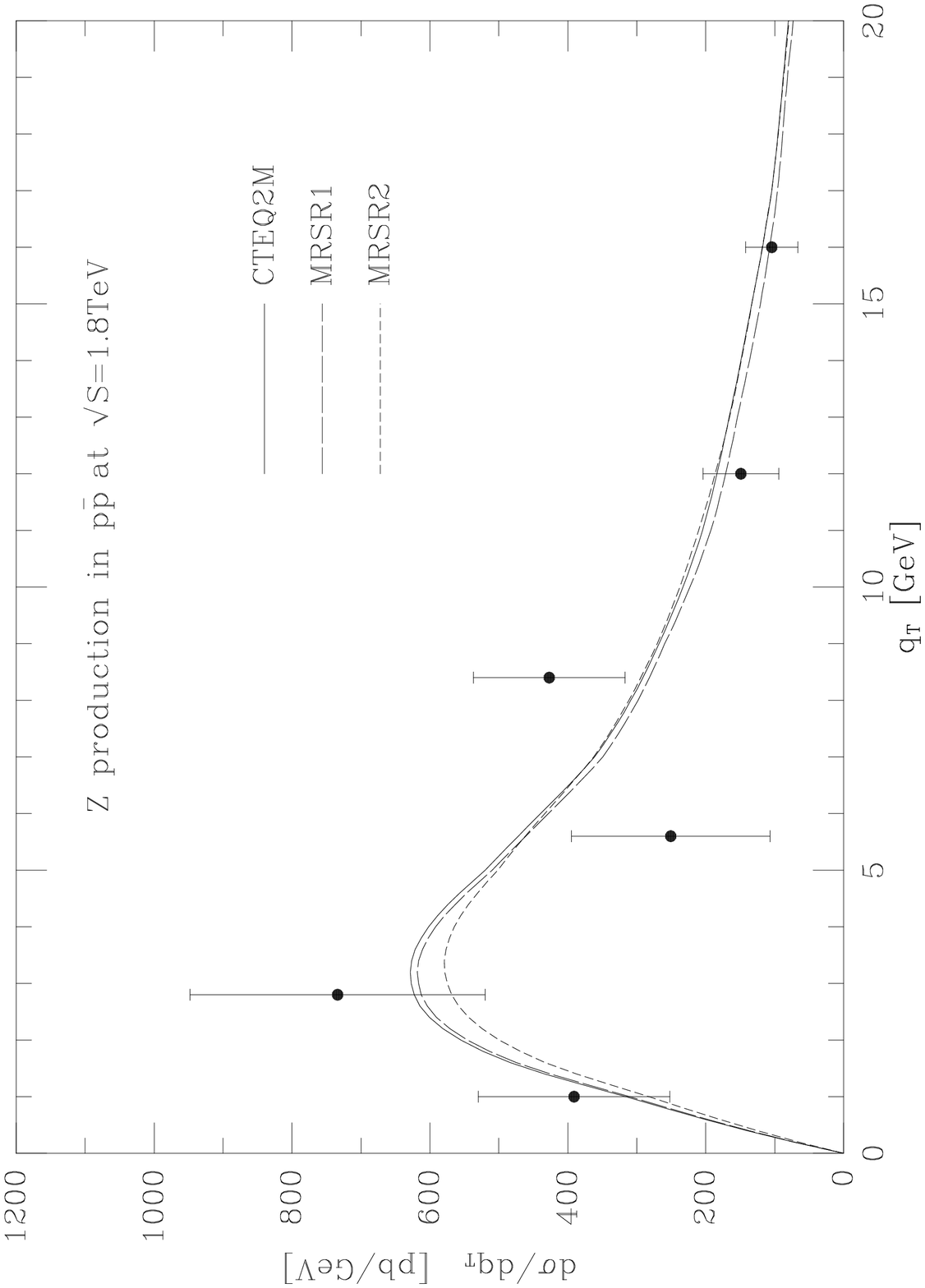}
\caption{Theoretical prediction for CDF $Z$ $d\sigma /d\protect \qt$.
These results are
obtained using several different parton distribution functions and with
an effective gaussian form of $F^{NP}$ ($g = 3.0\GeV^2,\blim=0.5\GeV^{-1}$).
}
\label{z}
\end{figure}

In Figures \ref{wpm} and \ref{z} we show our results for the
$W$ and $Z$ $\qt$ distributions. These results are obtained with $F^{NP}$
given in Eq. (\ref{gauss}), with $g=3.0\GeV^2$ and $\blim=0.5\GeV^{-1}$,
using several different parton distribution functions.
Besides illustrating the $\alpha_S$ dependence of
$d\sigma/d\qt$,\footnote{For CTEQ2M \protect\cite{CTEQ2},
MRSR1 and MRSR2 \protect\cite{MRS96}, we used
$\alpha_S(M_Z)$ of 0.110, 0.113 and 0.120, respectively.}
these two figures also show that the fit to the data is as good as the
one obtained in Ref. \cite{LY}, even though we used much simpler
functional form of $F^{NP}$.

\begin{figure}[p]
\vspace{12.0cm}
\includegraphics{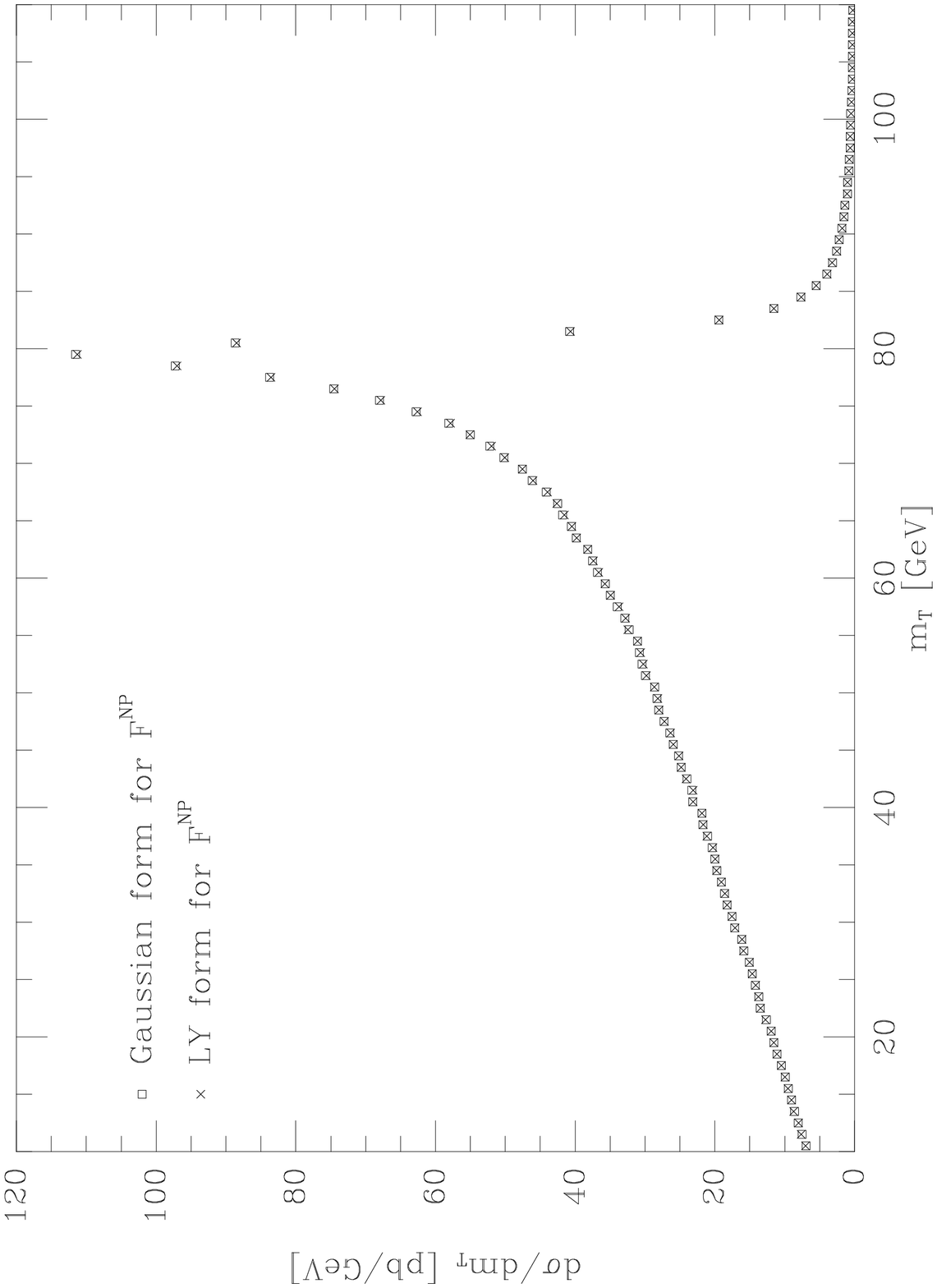}
\caption{Transverse mass distribution for
$W^++W^-$ production at Tevatron. We used
LY form (with their central values for the non-perturbative
parameters), and also an effective gaussian form of
$F^{NP}$ (with $g = 3.0\GeV^2$ and $\blim=0.5\GeV^{-1}$).
}
\label{transm}
\end{figure}

We now consider briefly the import of these results for the
measurement of the $W$ mass.
Figure \ref{transm} shows the transverse mass $m_T$ of
the lepton pair, obtained with the two different choices of $F^{NP}$.
To first order the transverse mass is insensitive to the
transverse motion of the $W$, and because of that
the  $m_T$ distribution is largely
independent of the non-perturbative parameters.

\begin{figure}[p]
\vspace{12.0cm}
\includegraphics{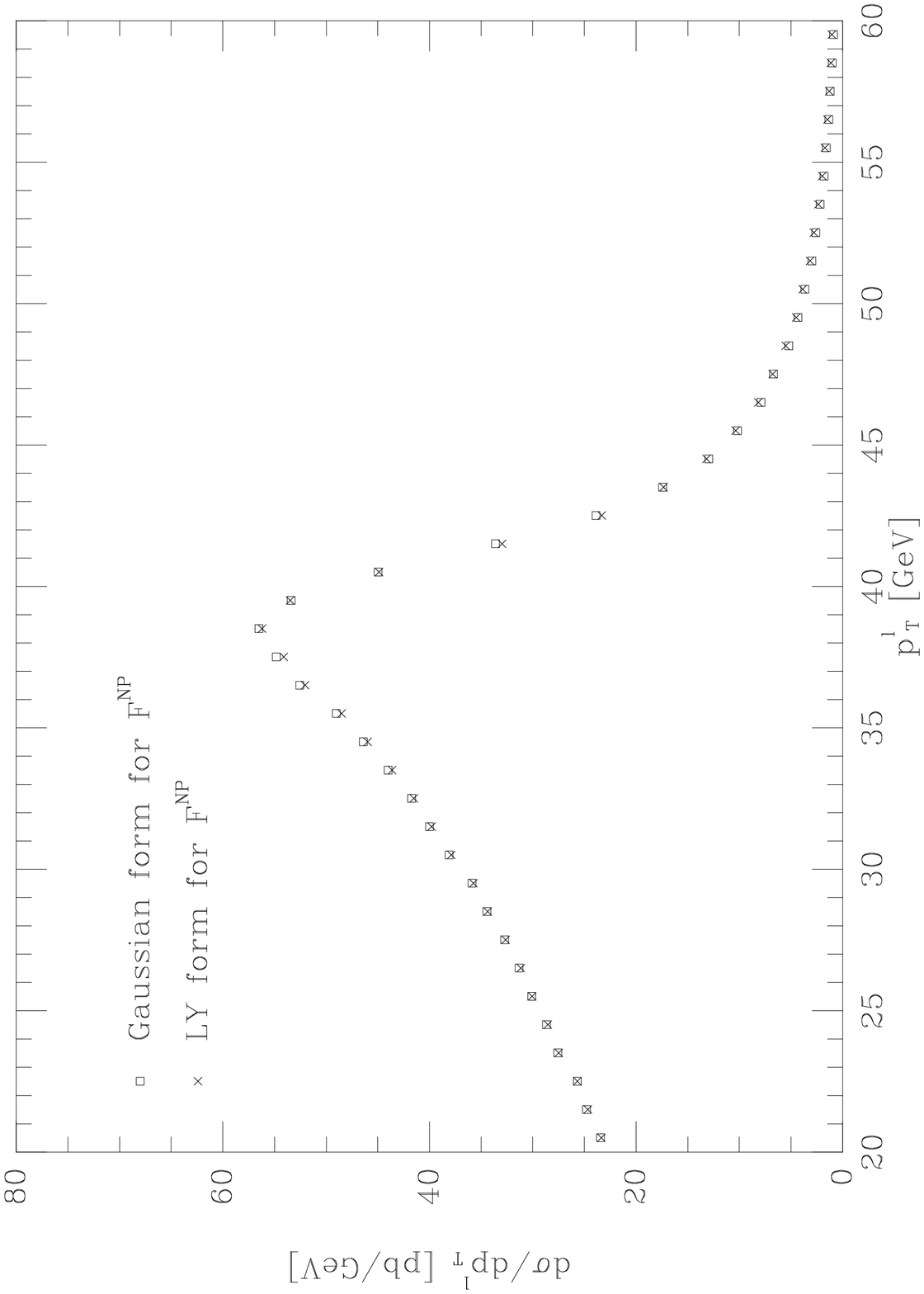}
\caption{$p_T^{l}$ distribution from
$W^-$ production at Tevatron. We used
LY form (with their central values for the non-perturbative
parameters), and also an effective gaussian form of
$F^{NP}$ (with $g = 3.0\GeV^2$ and $\blim=0.5\GeV^{-1}$).
}
\label{ptl}
\end{figure}

As the luminosity of the Tevatron is increased the number of interactions
per beam crossing will increase, leading to a degradation of the missing energy
resolution. Therefore the measurement of the Jacobian peak in the lepton
transverse momentum will become a competitive method of measuring the
$W$ mass. Figure \ref{ptl} shows the expected transverse momentum
distribution of the electron from $W^-$ decay. The width of this
distribution is broader than the transverse mass distribution
and the dependence on the non-perturbative functions is larger.
A quantitative estimate of the size of this dependence will have to
await a reliable extraction of the non-perturbative parameters.

\section{Conclusions}
\label{conc}

In view of the large number of $W$ and $Z$ bosons to be expected in
Run II at the Tevatron we have returned to consider
their production and decay in hadronic collisions.
We have provided a description of vector boson production which not
only  gives a correct description at small $\qt$, but also reproduces
the correct formula for the $\qt$-integrated cross section. In addition
we have included the decay of the vector bosons so that experimental
cuts can be included.

In the course of
our numerical work we have raised several issues which have not
been adequately addressed in the literature.
The analysis of the low energy experiments
needs to be repeated, using
the $\qt$-integrated data to fix the overall normalization, before any
attempt to determine the form of the non-perturbative function is made.
Furthermore, this analysis should include all low energy experiments
for which $\qt$-dependent distributions are available.  On the other
hand, for the $W$ mass measurement the effective form of
the non-perturbative function can
be extracted from the $Z$ data. This would eliminate uncertainties
related to the determination of $F^{NP}$ from the low energy experiments.

Besides the form of the non-perturbative function, there are also
other theoretical problems which have to be resolved.
In particular, we have shown the necessity for
extending the results of Ref. \cite{AK} to include  the vector boson
decay, if a more complete theoretical description of the leptons
coming from that decay is desired.
As shown in Ref. \cite{AK}, for the $W$ and $Z$ production at Tevatron
the $\cO(\alpha_S^2)$ calculation of the finite part of Eq. (\ref{totalcs})
should yield satisfactory results for matching of low and high
transverse momentum regions.
However, it is not quite clear whether such a calculation
would entirely solve the problem of getting a complete description
of the $\qt$ distribution for the $\gamma^*$ production
in the low energy experiments.

\begin{center}
ACKNOWLEDGMENTS
\end{center}
This work was supported in part by the U.S. Department of Energy
under Contracts No. DE-AC02-76CH03000.
R.K.E. and D.A.R. thank the CERN theory group for hospitality.

\appendix
\section{Details of the formulae}
\label{details}

\subsection{Couplings}

In this appendix we document the  results for functions $H^{(0)}$, $H^{(1)}$,
and $\Sigma^{(1)}$ which appear in Eqs.~(\ref{w}) and (\ref{finite2}).
These functions can all be separated into parts which are even and
odd under parity, e.g. we can write $H^{(0)}= H^{(0)+} + H^{(0)-}$.
Consequently, we first define ``plus'' and ``minus''
quark-quark and quark-gluon couplings as
\bea
\cV_{qq'}^+ = |V_{qq'}|^2 (\gl^2+\gr^2)(\fl^2+\fr^2)\ ,\\
\cV_{qq'}^- = |V_{qq'}|^2 (\gl^2-\gr^2)(\fl^2-\fr^2)\ ,
\eea
and
\bea
\cV_{qg}^+ = \sum_{q'}|V_{qq'}|^2 (\gl^2+\gr^2)(\fl^2+\fr^2)\ ,\\
\cV_{qg}^- = \sum_{q'}|V_{qq'}|^2 (\gl^2-\gr^2)(\fl^2-\fr^2)\ ,
\eea
where $\gl$, $\gr$, $\fl$ and $\fr$
are listed in Table \ref{couplings}. The coefficients $V_{qq'}$
are elements of the Cabibbo-Kobayashi-Maskawa matrix for the $W$
production, and are equal to $\delta_{qq'}$ in the case of $Z$
or massive photon $\gamma^*$.

In the case of $l^+ l^- $
production, which can proceed through the exchange of either a $Z$
or a $\gamma^*$, the above expressions need to be modified by
making the replacements ($Q_e = -1$, $Q_u=2/3$, and $Q_d=-1/3$)
\bea
(\gl^2+\gr^2)(\fl^2+\fr^2) & \rar &
(\gl^2+\gr^2)(\fl^2+\fr^2)  + 4 e^4 Q_f^2 Q_e^2 \chi_2(Q^2) \nonumber \\
&+& 2 e^2 Q_f Q_e (\gl \fl+\gr \fr +\gl \fr+\gr \fl) \chi_1(Q^2)\ ,\\
(\gl^2-\gr^2)(\fl^2-\fr^2)& \rar &
(\gl^2-\gr^2)(\fl^2-\fr^2) \nonumber \\
&+& 2 e^2 Q_f Q_e (\gl \fl+\gr \fr -\gl \fr-\gr \fl) \chi_1(Q^2)\ ,
\eea
with
\bea
\chi_1(Q^2) &=&\frac{(Q^2  -M_Z^2)}{Q^2} \ ,\\
\chi_2(Q^2) &=& \frac{(Q^2  -M_Z^2)^2+M_Z^2 \Gamma_Z^2} {Q^4} \ .
\eea
In these expressions $e^2$ is the electromagnetic charge which is taken to
run (down from its value at the $Z$ coupling)
using the one loop electromagnetic $\beta$ function,
\beq
\frac{d \alpha}{d \ln Q^2} = \frac{\alpha^2}{3 \pi}
\Big[\sum_{l}+3 \sum_{f} Q_f^2 \Big]\ .
\eeq
At $Q=5 \GeV$ we find that $\alpha\simeq 1/133$.

In our numerical work
we choose four input parameters\footnote{The
boson widths are fixed at their measured values
$$
\Gamma_W= 2.07 \GeV,\; \Gamma_Z= 2.49 \GeV\ .
$$
These, as well as all other parameters, are taken
from the Review of Particle Properties~\cite{PDG}.}
\bea
G_F &=&1.16639 \times 10^{-5}\GeV^{-2}\ ,  \nonumber \\
M_Z&=&91.187 \GeV\ ,   \nonumber \\
\alpha(M_Z)& =& (128.89)^{-1}\ ,\nonumber \\
M_W&=& 80.33 \GeV\ .
\eea
In terms of these parameters we can derive the $\rho$ parameter
\beq
\rho = \frac{M_W^2}{M_Z^2}
 \Bigg[1-\frac{\pi \alpha(M_Z)}{\sqrt{2} G_F M_W^2}
\Bigg]^{-1} = 1.00654\ ,
\eeq
which enters in the couplings (see Table \ref{couplings})
in the improved Born approximation:
\bea
\gw^2 &=& 4 \sqrt{2} M_W^2 G_F  \ ,   \nonumber \\
\gz^2 & =& \sqrt{2} G_F M_Z^2 \rho \ ,\nonumber \\
x_W &=& 1-\frac{M_W^2}{\rho M_Z^2}\ .
\eea

\subsection{Matrix elements}

We start by considering the lowest order process for
the production of a vector boson of mass $\MB$ and width $\GB$
(the momenta are shown in brackets),
\beq
q(\pu) +\bq(\pd) \to l(\ku) + \bl (\kd)\ .
\eeq

The matrix element squared for lowest order process averaged (summed)
over the initial (final) spins and colours $(N=3)$ is given by
\beq
\overline{\sum} |M_{q\bq}|^2 = \frac{1}{4N}
 \frac{Q^4}{(Q^2  -\MB^2)^2+\MB^2 \GB^2} H^{(0)}_{q\bq} \ ,
\eeq
where
\beq
\hspace*{-2mm}
H^{(0)}_{q\bq} =
\frac{8 }{Q^4}
\left[
  \cV_{q\bq}^+ \; (\pu\cdot\kd \; \pd\cdot\ku +\pu\cdot\ku \; \pd \cdot \kd )
+ \cV_{q\bq}^- \; (\pu\cdot\kd \; \pd\cdot\ku -\pu\cdot\ku \; \pd \cdot \kd )
\right] \ .
\label{lowestorder}
\eeq

It is convenient to express the above matrix element in
terms of angular variables in the Collins-Soper (CS) frame, which is
defined by
\bea
\ku^\mu &=&\frac{Q}{2}
\left(1,\sin\phi \sin\theta,\cos\phi \sin\theta,\cos\theta\right)
\ ,\nonumber\\
\kd^\mu &=&\frac{Q}{2}
\left(1,-\sin\phi \sin\theta,-\cos\phi \sin\theta,-\cos\theta\right)
\ ,\nonumber\\
\pu^\mu &=&-\frac{(t-Q^2)}{2 Q} \left(1,0,-\sin\beta,\cos\beta\right)
\ ,\nonumber\\
\pd^\mu &=&-\frac{(u-Q^2)}{2 Q} \left(1,0,-\sin\beta,-\cos\beta\right)\ ,
\label{CSdefn}
\eea
with\footnote{In lowest order the
vector boson is produced at zero transverse momentum
and hence $\beta=0$.}
\beq
\tan \beta = \frac{\qt}{Q}\ .
\eeq
In the terms of  the vector boson  transverse momentum ($\qt=|\bqt|$)
and rapidity ($y$) in the lab frame, we also have
\beq
q^\mu =(M_T \cosh y, \bqt, M_T \sinh y)\ ,
\eeq
where $M_T^2=Q^2+\qt^2$ and $Q^2 = q^2$.
We further define several functions which determine the
angular dependence as
\bea
\lz &=& 1 + \cos^2{\theta}\ ,\nonumber \\
\az &=& {1 \over 2} (1 - 3 \cos^2{\theta})\ , \nonumber  \\
\au &=& \sin{2 \theta} \cos{\phi}\ , \nonumber \\
\ad &=& {1 \over 2} \sin^2{\theta} \cos{2 \phi}\ ,\nonumber \\
\at &=& 2 \cos{\theta}\ ,\nonumber  \\
\aq &=& \sin{\theta} \cos{\phi}\ .
\eea
Using these angular variables
Eq.~(\ref{lowestorder}) may be written
as,
\beq
 H^{(0)}_{q\bq}(\theta)=  \cV^{+}_{q\bq}\lz  + \cV^{-}_{q\bq} \at \ .
\label{h0}
\eeq
Note the relation between the $q\bar{q}$ and $\bar{q}q$ processes,
\beq
H^{(0)}_{\bq q}(\theta)=  H^{(0)}_{q\bq}(\pi-\theta)\ .
\eeq

We next consider $\ovl{H}^{(1)}$ and  $\Sigma^{(1)}$
(needed for the finite part of Eq.~(\ref{totalcs})), which are derived from
the matrix element squared for processes involving one parton emission,
and from the pieces removed from the cross section in the
resummation procedure, ({\it cf.}~Eqs.~(\ref{finite1}) and (\ref{finite2})).
It is sufficient to calculate results for the two-to-three processes
\bea
q(\pu) +\bq(\pd) &\to& l(\ku) + \bl(\kd) +g(\kt)\ , \\
\label{qgllbq}
q(\pu) +g(\pd) &\to& l(\ku) + \bl (\kd) +q'(\kt)\ ,
\label{qqbllbk}
\eea
as all other processes are determined by the crossing relations.
The invariant variables for the above processes are
\bea
s &=&  (\pu+\pd)^2\ , \nonumber \\
q^2&=&Q^2 =(\ku+\kd)^2\ , \nonumber \\
t &=&  (\pu-q)^2 =(\pd-\kt)^2\ , \nonumber \\
u &=&  (\pd-q)^2=(p-\kt)^2\ .
\eea
The matrix element squared can be put in the form
\beq
\overline{\sum} |M_{ab}|^2 = {g^2 \over 2 N}
\frac{Q^2}{(Q^2  -\MB^2)^2+\MB^2 \GB^2} H^{(1)}_{ab} \ ,
\eeq
with
\bea
H^{(1)}_{q\bq} &=& \frac{8 C_F }{t u}
\Bigg\{\cV^+_{q\bq}
\left[(\pd \cdot\ku)^2+(\pu\cdot\kd)^2 + (\pd \cdot\kd)^2+(\pu\cdot\ku)^2
\right]
\nonumber \\
&+& \cV^-_{q\bq}
\left[(\pd \cdot\ku)^2+(\pu\cdot\kd)^2 -(\pd \cdot\kd)^2-(\pu\cdot\ku)^2
\right] \Bigg\} \ ,
\label{resforjk}\\
H^{(1)}_{qg} &=& -\frac{8 T_R}{t s}
\Bigg\{\cV^+_{qg}
\left[(\kt \cdot\ku)^2+(\pu\cdot\kd)^2
 +(\kt \cdot\kd)^2+(\pu\cdot\ku)^2 \right]
\nonumber \\
&+& \cV^-_{qg}
\left[(\kt \cdot\ku)^2+(\pu\cdot\kd)^2
 -(\kt \cdot\kd)^2-(\pu\cdot\ku)^2 \right] \Bigg\} \ .
\label{resforjg}
\eea
Again, using Eq.~(\ref{CSdefn}) the functions $H^{(1)}$ can be expressed
in terms of invariant variables and the  angles in the CS frame
\beq
H^{(1)}_{ab} \to  H^{(1)}_{ab} (Q^2,t,u,\theta,\phi) \ .
\eeq

Before writing down the expressions
for $\ovl{H}^{(1)}$ and  $\Sigma^{(1)}$,
we clarify the relationship
between different variables. The Mandelstam variables for the parton
subprocess can be expressed in
terms of the integration variables in Eq.~(\ref{finite})
($\za=\xa/\xia $ and $\zb=\xb/\xib $) as
\bea
s&=&\frac{Q^2}{\za \zb}\ ,\nonumber \\
t&=&-\frac{Q^2}{\za}(\eta -\za)\ ,\nonumber \\
u&=&-\frac{Q^2}{\zb}(\eta -\zb)\ ,
\eea
where $M_T = \sqrt{Q^2+\qt^2}$ and
$\eta = M_T/Q$. We further note that
if $Q^2 = s+t+u$ and $\qt^2 = ut/s$ one can
derive
\beq
\eta =\frac{M_T}{Q}\equiv \frac{1+\za\zb}{\za+\zb}\ ,
\eeq
so that $Q^2 (\equiv \qt^2/(\eta^2-1)$) can be expressed in terms of $\qt^2$,
$\za$ and $\zb$.

For the process $q\bq\rar l\bl  g$
we have $\ovl{H}^{(1)}_{q\bq}= \ovl{H}^{(1)+}_{q\bq} + \ovl{H}^{(1)-}_{q\bq}$,
with\footnote{The term proportional to $\au$ in Eq.~(\ref{Hqq+})
differs from the analogous expression in Eq.~(16) of Ref.~\cite{BQY}.
In addition, Eqs.~(16)-(18) of Ref.~\cite{BQY} are missing colour factors.}
\bea
\ovl{H}^{(1)+}_{q \bq}(Q^2 ,t,u,\theta,\phi)
&\hspace*{-0.5mm}=\hspace*{-0.5mm}&
\cV^+_{q\bq}\frac{C_F}{s} \Bigg\{
\RP(t,u)  {1 \over M_T^2} (\lz + \az + \ad )
- \RM(t,u)   {Q \over \qt  M_T^2} \au \Bigg\}  \ ,\hspace*{+8mm}
\label{Hqq+}\\
\ovl{H}^{(1)-}_{q \bq}(Q^2 ,t,u,\theta,\phi)
&\hspace*{-0.5mm}=\hspace*{-0.5mm}&
\cV^-_{q\bq}\frac{C_F}{s} \Bigg\{
\RP(t,u) \frac{Q}{\qt ^2 M_T} \Big(1-\frac{Q}{M_T}\Big) \at
-\RM(t,u) {2 \over \qt  M_T} \aq \Bigg\}\ .
\label{h1pqbq}
\eea
Here,
\beq
\cR_{\pm}(t,u) = (Q^2 - t)^2 \pm (Q^2 - u)^2\ .
\eeq
Similarly, for $qg \rar l\bl q'$
($\ovl{H}^{(1)}_{qg}= \ovl{H}^{(1)+}_{qg} + \ovl{H}^{(1)-}_{qg}$) we have
\bea
\ovl{H}^{(1)+}_{qg}(Q^2 ,t,u,\theta,\phi) &=&
-\cV^+_{qg}\frac{T_R}{st} \Bigg\{ \Big( \RP(t,s) - \RP(0,s) \Big) \lz
+ {\qt ^2 \over M_T^2} \Bigg[ \RP(t,-s) \left( \az + \ad \right)
\nonumber \\
&-&\left( (Q^2 - t)^2 + \RM(t,u) \right) {Q \over \qt }   \au \Bigg] \Bigg\}
\ ,\\
\ovl{H}^{(1)-}_{qg}(Q^2 ,t,u,\theta,\phi) &=&
-\cV^-_{qg}\frac{T_R}{st} \Bigg\{
\Big( {Q \over M_T} \RP(t,s)-\RP(0,s)
- 2 {Q \over M_T} t (Q^2 - s) \Big) \at
\nonumber \\
&-& {2 \qt \over M_T} \left[ 2 s (Q^2-s) +\RP(t,s)\right]\aq \Bigg\}\ .
\label{h1mqg}
\eea
Note that the crossing relationships for all other two-to-three processes are
given as
\bea
\ovl{H}^{(1)}_{\bq q}(Q^2 ,t,u,\theta,\phi) &=&
\ovl{H}^{(1)}_{q\bq}(Q^2 ,u,t,\pi-\theta,\phi)\ ,
\nonumber\\
\ovl{H}^{(1)}_{gq}(Q^2 ,t,u,\theta,\phi) &=&
\ovl{H}^{(1)}_{qg}(Q^2 ,u,t,\pi-\theta,\phi) \ ,
\nonumber\\
\ovl{H}^{(1)}_{\bq g}(Q^2 ,t,u,\theta,\phi) &=&
\ovl{H}^{(1)}_{qg}(Q^2 ,t,u,\pi-\theta,\pi-\phi) \ ,
\nonumber\\
\ovl{H}^{(1)}_{g\bq}(Q^2 ,t,u,\theta,\phi) &=&
\ovl{H}^{(1)}_{qg}(Q^2 ,u,t,\theta,\pi-\phi) \ .
\label{cross}
\eea

The corresponding expressions for
$\Sigma^{(1)}_{q\bq}$ and $\Sigma^{(1)}_{qg}$
look simpler  when written in terms of $\za$ and $\zb$,
\bea
\Sigma^{(1)}_{q \bq} (Q^2 ,\za,\zb,\theta) &=&
 r_{q \bq}(Q^2,\za,\zb) {Q^2 \over \qt^2}
(\cV^+_{q\bq} \lz  +\cV^-_{q\bq}\at) \ ,
\label{s1p}\\
\Sigma^{(1)}_{q g}(Q^2 ,\za,\zb,\theta) &=&
r_{q g}(Q^2 ,\za,\zb) {Q^2 \over \qt^2}
(\cV^+_{qg} \lz + \cV^-_{qg}\at)\ .
\label{s1m}
\eea
Here, the functions functions $r$ are given as
\bea
r_{q\bq}(Q^2,\za,\zb)& =& C_F
\Bigg\{  (\za^2+\zb^2)
 \delta (1-\eta \za -\eta \zb+\za \zb)
  \nonumber \\
&-&
 \Theta(Q^2-q_T^2) \
 \delta (1 - \zb) \left( {1 + \za^2 \over 1 - \za} \right)_+
 -  \Theta(Q^2-q_T^2) \
 \delta (1 - \za) \left( {1 + \zb^2 \over 1 - \zb} \right)_+
  \nonumber \\
& - &2  \Theta(Q^2-q_T^2) \ \delta (1 - \za)  \delta (1 - \zb)
      \left[ \ln{\left( {Q^2 \over q_T^2} \right)} - {3 \over 2}\right]
  \Bigg\} \ ,
\label{rjk}
\eea
and
\bea
 r_{qg}(Q^2,\za,\zb) &=& T_R
\Big\{ \za (\eta-\zb)  \left[\za^2 \zb^2 +(1-\za \zb)^2 \right]
 \delta (1-\eta \za -\eta \zb+\za \zb)
\nonumber \\
&-&  \Theta(Q^2-q_T^2) \
  \left[\zb^2 +(1-\zb)^2 \right] \delta (1 - \za)  \Big\}\ .
\label{rqg}
\eea
When using crossing relationships
analogous to Eq.~(\ref{cross}) for $\Sigma^{(1)}$
one should remember that $t\lrr u$ implies $\za \lrr \zb$.

 From Eqs.~(\ref{h1pqbq})-(\ref{h1mqg}) it is evident that $\ovl{H}^{(1)}$ is
an integrable function of $q_T^2$ at $\qt^2\rar 0$.
We now show this is also true of
Eqs.~(\ref{s1p}) and (\ref{s1m}).
Define
\beq
\xas = \frac{1-\eta \xb}{\eta-\xb}\ ,\
\xbs = \frac{1-\eta \xa}{\eta-\xa}\ ,
\eeq
and consider the integral $J_{ab}$
\beq
J_{ab}
= \int_{\xa}^1 \frac{d \za}{\za}
\int_{\xb}^1 \frac{d \zb}{\zb}\;
 f_a\Big(\frac{\xa}{\za}\Big) f_b\Big(\frac{\xb}{\zb}\Big)\;
  r_{ab}(Q^2,\za,\zb)\ .
\eeq
Inserting the explicit form of $r_{q \bq}$
from Eq.~(\ref{rjk}) and using the representation
\beq
\ln \frac{\qt^2}{Q^2}\equiv \ln [\eta^2-1]= \ln\big[(\eta-\xa)(\eta-\xb)\big]
-\int_{\xa}^{\xas} \frac{d\za}{\eta-\za} \ ,
\eeq
we find
\bea
\hspace*{-5mm} J_{q\bq} & = &  \int_{\xa}^{1} d \za
\int_{\xb}^1 d \zb  \delta (1-\eta \za -\eta \zb+\za \zb) \nonumber \\
&\times &
\bigg[g_{q\bq}(\za,\zb)-g_{q\bq}(\za,1)-g_{q\bq}(1,\zb)+g_{q\bq}(1,1)\bigg]
  \nonumber \\
&+&
\int_{\xa}^{1} d \za \bigg[ g_{q\bq}(\za,1)-g_{q\bq}(1,1)\bigg]
\bigg[ \frac{\Theta(\xas-\za)}{(\eta-\za)}
       -\frac{ \Theta(Q^2-q_T^2) }{(1-\za)} \bigg]
  \nonumber \\
&+&\int_{\xb}^1 d \zb \bigg[ g_{q\bq}(1,\zb) -g_{q\bq}(1,1) \bigg]
\bigg[ \frac{\Theta(\xbs-\zb)}{(\eta-\zb)}
 -\frac{ \Theta(Q^2-q_T^2) }{(1-\zb)} \bigg]
  \nonumber \\
&+&  g_{q\bq}(1,1)
\bigg[
\ln \bigg( \frac{(\eta-\xa)(\eta-\xb)}{(1-\xa)(1-\xb)} \bigg)
+ \Theta(q_T^2 - Q^2) \
\ln \bigg( \frac{(1-\xa)(1-\xb)}{\eta^2-1} \bigg) \bigg]\ ,
\eea
with
\beq
g_{q\bq}(\za,\zb)= C_F \frac{\za^2+\zb^2}{\za \zb}
 f_q\Big(\frac{\xa}{\za}\Big) f_{\bq}\Big(\frac{\xb}{\zb}\Big)\ .
\eeq
In this form the vanishing of $J_{q\bq}$
in the limit $\eta \to 1$ is manifest. Similarly, using Eq.~(\ref{rqg})
for $r_{qg}$ we obtain
\beq
J_{qg}= \int_{\xb}^{1} d\zb \Big[g_{qg}(\zas,\zb)\Theta(\xbs-\zb)
-\Theta(Q^2-\qt^2) g_{qg}(1,\zb)\Big]\ ,
\eeq
where $\zas = (1-\eta \zb)/(\eta-\zb) $ and
\beq
g_{qg}(\za,\zb) = T_R \frac{\za^2 \zb^2 +(1-\za\zb)^2}{\zb}
 f_q\Big(\frac{\xa}{\za}\Big) f_g\Big(\frac{\xb}{\zb}\Big)\ .
\eeq
Clearly, in the limit $\eta \rar 1$ we have that $J_{qg}\rar 0$.

\subsection{``Plus'' distributions}

The definition of single plus distribution used
in Section \ref{ics} and in this appendix  is the standard one,
\beq
\int_x^1 dz\ f(z) \left[g(z)\right]_+=
\int_x^1 dz\ [f(z)-f(1)]g(z) -f(1)\int_0^x dz\ g(z)\ .
\eeq
In Section \ref{ics} we also used the double plus distribution
defined as
\bea
&&  \hspace*{-1.3cm}
\int_{\xa}^1 d\za
\int_{\xb}^1 d\zb f(\za,\zb)
\left(\frac{1} {(1-\za)(1-\zb)}\right)_{++}=\nonumber \\
&=&
\int_{\xa}^1 d\za
\int_{\xb}^1 d\zb
\frac{f(\za,\zb)-f(\za,1)-f(1,\zb)+f(1,1)}{(1-\za)(1-\zb)}
\nonumber \\
&+& \int_{\xa}^1 d\za f(\za,1)\left(\frac{1}{1-\za}\right)_+ \ln(1-\xb)
 + \int_{\xb}^1 d\zb f(1,\zb)\left(\frac{1}{1-\zb}\right)_+  \ln(1-\xa)
\nonumber \\
&-&f(1,1) \Bigl. \ln(1-\xa)   \ln(1-\xb) \Bigr. \ .
\eea

\section{Numerical evaluation of Bessel transform}

When $q$ and $b$ become large it is convenient to use the asymptotic
expansion\footnote{This method was suggested in \cite{AK}.
The formula  corresponding to Eq.~(\ref{AKformula})
in Ref.~\cite{AK} contains an overall
sign error.}
to evaluate that portion of the Bessel transform
because of the
large cancellations between different cycles of the Bessel function.
For large $b$ we find that
\bea \label{AKformula}
 \int_b^\infty dy \; J_0(qy ) h(y) &=&
 \sqrt{\frac{2}{\pi q}} \sum_{n=0}^\infty \frac{1}{q^{n+1}}
\sum_{k=0}^{n} \frac{\Gamma(\frac{1}{2}+k)}{2^k k! \Gamma(\frac{1}{2}-k)}
\nonumber \\
&\times & \cos\Big(qb + n \frac{\pi}{2}+\frac{\pi}{4}\Big)
\; \frac{d ^{(n-k)}}{d^{(n-k)} b}
\Bigg[ \frac{h(b)}{b^{k+\frac{1}{2}}}\Bigg]\ .
\eea
If we further make the special choice of $b=b_s$
\beq
b_s=\Big(2 s+\frac{1}{4}\Big)\frac{\pi}{q}\ , \;\; s =1,2,3 \ldots\ ,
\eeq
the even $n$ terms will cancel because
\beq
\cos\Big(qb_s + n \frac{\pi}{2}+\frac{\pi}{4}\Big)
 = -\sin\Big( n \frac{\pi}{2}\Big)\ ,
\eeq
and we have
\bea \label{bessexp}
&& \int_{b_s}^\infty dy \; J_0(q y) h(y) =
 -\frac{1}{q} \sqrt{\frac{2}{\pi}}
 \frac{1}{(qb_s)^\frac{3}{2}} \Bigg\{
  \Big[ b_s h^\prime(b_s) -\frac{5}{8} h(b_s) \Big]
  + O\Big(\frac{1}{(qb_s)^{2}}\Big) \Bigg\} \ .
\eea
In practice it is found that the explicit integration needs to be done
over only a few cycles, before approximating with the asymptotic expansion,
Eq.~(\ref{bessexp}).

\clearpage

\renewcommand\arraystretch{1.3}

\begin{table}
\caption{The range of values for $g$ for which results
shown in Figures \protect \ref{blim_r209} through \protect \ref{blim_e605}
were obtained.
}
\begin{center}
\begin{tabular}{|c|c|c|c|}
\hline
Experiment & Range in $Q\ [\GeV]$ & Range in $\sqrt{\tau}$
& Range in $g\ [\GeV^2]$
\\
\hline
R209  &  5 -- 8  & 0.081 -- 0.129    & 0.52 -- 0.55  \\
\hline
E288  &  5 -- 6  & 0.182 -- 0.219    & 0.30 -- 0.35 \\
  &  6 -- 7      & 0.219 -- 0.255  &0.32 -- 0.40 \\
  &  7 -- 8      & 0.255 -- 0.292  &0.33 -- 0.46 \\
  &  8 -- 9      & 0.292 -- 0.328   &0.38 -- 0.50 \\
\hline
E605 &  7 -- 8   & 0.180 -- 0.206  &0.34 -- 0.41 \\
 &  8 -- 9       & 0.206 -- 0.232  &0.42 -- 0.49 \\
 &  10.5 -- 11.5 & 0.271 -- 0.296 &0.46 -- 0.53 \\
 &  11.5 -- 13.5 & 0.296 -- 0.348 &0.46 -- 0.51 \\
\hline
\end{tabular}
\end{center}
\label{g_range}
\end{table}
\begin{table}
\caption{Vector boson couplings in the notation given in
the Appendix \protect\ref{details}. Lepton and quark charges
are $Q_e = -1$, $Q_u=2/3$, and $Q_d=-1/3$.}
\begin{center}
\begin{tabular}{|c|c|c|c|c|}
\hline
Boson & $\fl$ & $\fr$ & $\gl$ & $\gr $  \\
\hline
$W$  & $\gw/\sqrt{2}$ & $ 0 $ & $\gw/\sqrt{2}$ & $0$ \\
\hline
$Z$  & $\gz  (T_3-Q_e \xw ) $ &
$-\gz  Q_e \xw  $ &
$\gz  (T_3-Q_f \xw ) $ &
$-\gz  Q_f \xw  $ \\
\hline
$\gamma^*$ &$ e Q_e $ & $ e Q_e $ & $ e Q_f $ & $ e Q_f $ \\
\hline
\end{tabular}
\end{center}
\label{couplings}
\end{table}

\renewcommand\arraystretch{1.}

\clearpage
\newpage

\end{document}